\documentclass[usenatbib]{mn2e}

\usepackage[dvips]{graphicx}
\usepackage{rotating}
\usepackage{color}

\begin{document}

\def\HI{H{\sc i} }
\def\HII{H{\sc ii} }
\def\OVI{O{\sc vi}}
\def\Ha{{\rm H}\alpha }
\def\Msun{M_{\odot} }
\def\kms{\,km\,s^{-1}}
\def\Msunrm{{\mathrm M}_{\odot} }
\def\Lsunrm{{\mathrm L}_{\odot} }

\title{Chemo-dynamical evolution of tidal dwarf galaxies. \\ I. Method and IMF dependence}

\author[S. Ploeckinger et al.]
	{S.~Ploeckinger,$^1$\thanks{email: sylvia.ploeckinger@univie.ac.at} G.~Hensler,$^1$  S.~Recchi,$^1$ N.~Mitchell,$^1$ P.~Kroupa$^2$ \\
	$^1$University of Vienna, Department of Astrophysics, T\"urkenschanzstr. 17, 1180 Vienna, Austria \\
	$^2$ Helmholtz-Institut f\"ur Strahlen- und Kernphysik, Nussallee 14-16, 53115 Bonn, Germany}

\maketitle

\begin{abstract}
We present high-resolution simulations of tidal dwarf galaxies (TDG) to investigate their early chemo-dynamical evolution and test their survivability. In this work the simulation setup is introduced and the response of TDGs to self-consistent star formation (SF) and an external tidal field is examined. Throughout the simulation star cluster particles with variable masses down to $5\,\Msunrm$ form, depending on the local gas reservoir.  For low cluster masses $M_{\mathrm{cl}}$, the stellar initial mass function (IMF) is considered to be either filled or truncated at a maximal star mass $m_\mathrm{max}$ to represent the observed $m_{\mathrm{max}} - M_{\mathrm{cl}}$ relation (IGIMF theory). The evolution of TDGs with fully-populated and truncated IMFs are compared to study the impact of stellar energy feedback on their survivability. Both TDGs experience an initial starburst but after a dynamical time they evolve into dwarf galaxies with self-regulated and continuous SF. At this stage the truncated-IMF model contains about 6 times more stellar mass than the invariant IMF models, but the final bound gas mass is comparable in both models. In spite of their significantly different SF histories, both TDG models are not disrupted within the first 500 Myr. We conclude that TDGs can survive an early starburst, independent of the underlying IMF description, even though they do not harbor a stabilizing dark matter halo.
\end{abstract}


\section{Introduction}
Long filamentary structures, either connected to perturbed galaxies or building bridges between an interacting pair of galaxies, were first observed and categorized as peculiar objects in the \citet{1966ApJS...14....1A} catalogue. \citet{1972ApJ...178..623T} performed the first restricted N-body simulations of galaxies on close encounters and found that  the extended structures that appear during the interaction are caused by tidal forces. Since that time detailed observational data of interacting galaxies with tidal arms starting with \citet{1956ErNW...29..344Z} and \citet{1994AA...289...83D} has been collected and analyzed, showing that many tidal arms contain kinematically decoupled structures with active star formation in the same mass range as dwarf galaxies, the so-called tidal dwarf galaxies (TDG, see also \citet{2013MNRAS.429.1858D} for a review). Because of their top-down formation scenario they cannot contain dark matter (DM), as emphasized by \citet{2007AA...472L..25G}, \citet{2010AdAst2010E...1B}, \citet{2012PASA...29..395K} and \citet{2013MNRAS.429.1858D}. 

Since TDGs are formed from gas already enriched by stars from the more massive progenitor galaxies, they deviate from the metallicity-luminosity relation \citep{2004ApJ...613..898T}, meaning that their metallicity is higher than in classical dwarf galaxies with similar luminosities \citep[i.~e.][]{2003AA...397..545W}. Depending on their formation time the deviation in the luminosity-metallicity relation varies and TDGs formed at high redshifts may be as metal-poor as classical dwarf galaxies. 
In the star forming regions within the gaseous bridge (Arp's loop) between M81 and M82, \citet{2008AJ....135..548D} found in addition to the young stellar population formed in the tidal debris \citep{2002AA...396..473M}, an old ($>$ 1 Gyr) stellar population, which may have formed in the stellar disks of the host galaxies and ejected during the passage. In other galaxy groups, as in Stephan's Quintet the analyzed spectral energy distribution does not show evidence for an old stellar population with ages higher than 1 Gyr \citep{2010AJ....140.2124B}.

Starting with the very basic experiments of galaxy interactions by \citet{1941ApJ....94..385H}, numerical simulations with different state-of-the-art numerical techniques, such as pure N-body, Smoothed Particle Hydrodynamic (SPH) or Adaptive mesh refinement (AMR) have been performed in the last decades with increasing resolution and have given constraints to a possible formation scenario of TDGs. \citet{2007MNRAS.375..805W} found that dissipative processes play an important role in the initial collapse of part of the tidal tail into a bound structure. While their low resolution (288k particles) N-body models reproduce the observed clumpiness, larger particle numbers (up to 4128k particles), i.e. higher mass resolution, smooth out the tails. Only when dissipation is included by an SPH treatment, the clumpiness occurs again in their models. \citet{2009ApJ...706...67R} investigated the importance of tidal modes, which develop during galaxy interactions where two potentials overlap. They found that regions with fully compressive tides occur frequently and can trigger the formation or prevent the disruption of star clusters or TDGs. \citet{2006AA...456..481B} calculated the long-term evolution of TDGs with SPH and found that TDGs which were formed at the tips of the tidal arm can survive for more than 2 Gyr. \citet{2007AA...470L...5R} for the first time performed 2D hydrodynamical simulations to study the chemo-dynamical evolution of young TDGs for the first 300 Myr while \citet{1997NewA....2..139K} and \citet{2012MNRAS.424.1941C} investigated the long term evolution and survivability of TDGs with stellar dynamics.
The results of this research demonstrate that TDGs can become long-lived satellite galaxies comparable to the known MW satellites \citep[see also][]{2007MNRAS.376..387M}.

As TDGs reach between a few times $10$ and $100$ kpc into the galactic environment, they probe the DM distribution of the outer halo. \citet{2003AA...411L.469B} showed with N-body simulations that extended dark matter haloes, up to 10 times the radius of the stellar disk, are needed to form massive, of the order of  $10^9\, \Msunrm$ gas accumulations at the tip of the tidal arms. Simulations of truncated haloes (at 3 times the optical radius of the galaxy) with otherwise the same initial conditions lead to several smaller ($10^7 \dots 10^8\,\Msunrm$) gravitationally collapsing clumps along the tidal tails, but without a significant gas accumulation at the tip of the arm. \citet{2004AA...427..803D} presented a toy model to show that in truncated haloes the tidal arms would be stretched and therefore fragmentation and a subsequent collapse into structures in the mass range of dwarf galaxies is not possible. 

Statistical analysis of observations and simulations of interacting galaxies and the number of produced TDGs illustrates that a significant fraction of today's dwarf galaxy population could have had a tidal origin. The estimates are based on both the production rate per merger and the survival timescale of TDGs and span from a few percent f.e. by  \citet{2006AA...456..481B} (10 \%) and by \citet{2012MNRAS.419...70K} (6 \%)  where only mergers in the local universe were considered, to up to 100\% \citep{2000ApJ...543..149O} if on average 1-2 long-lived TDGs are produced per merger throughout the whole galaxy formation time span of the universe. The recent findings about the 3D structure of the Local Group (LG), including a vast polar structure (VPOS) of satellite galaxies, globular clusters and stellar streams \citep{2012MNRAS.423.1109P} surrounding the Milky Way Galaxy (MWG) and a vast thin disk of satellites (VTDS), a similar but even more extreme structure around M31 \citep{2013Natur.493...62I} emphasize the importance of understanding the evolution of TDGs.

\paragraph*{Initial mass function (IMF):}
  The stellar IMF has been first introduced by \citet{1955ApJ...121..161S} as a
convenient way of parameterizing the relative number of stars as a
function of their mass.  Since then, the IMF has been shown to be a
fundamental distribution function which governs the evolution of galaxies \citep{2001MNRAS.322..231K, 2002Sci...295...82K, 2003PASP..115..763C, 
2010ARAA..48..339B, 2006ApJ...648..572E, 2011ApJ...731...61E, 2013pss5.book..115K, 
2013MNRAS.434...84W}. It is well described by two power laws \citep{2001MNRAS.322..231K} 
or by a log-normal distribution plus power-law extensions 
\citep{2003PASP..115..763C} with a similar shape.  

Various studies \citep[i.~e.][]{2002Sci...295...82K, 2003ARAA..41...15M, 2010ARAA..48..339B}
suggest that the IMF is universal, meaning that its shape is the same
in various stellar systems both in the Milky Way and in the Magellanic
clouds. \citet{2009ApJ...706..599L} measured the H$\alpha$ flux, a tracer for the most massive O and early-type B stars ($M_* \ge 17\,\Msunrm$), and the far ultraviolet (FUV) flux, a tracer for stars with $M_* \ge 3\,\Msunrm$, in a sample of $\approx 300$ star-forming galaxies within 11 Mpc of the Milky Way. 
They found H$\alpha$-to-FUV flux ratios up to an order of magnitude lower than expected in dwarf galaxies with star formation rates (SFR) of $\mathrm{SFR} \le 0.1\,\Msunrm\,\mathrm{yr}^{-1}$. They conclude that an IMF that is deficient in high-mass stars for dwarf and low surface brightness galaxies is consistent with their data. In addition, it is clear that there must be some correlation
between the mass of a star cluster and the uppermost stellar mass within
each cluster, for the simple reason that it is very unlikely to form
the most massive stars in very low-mass clusters.  This leads to the so-called
maximum-mass vs. embedded cluster mass ($m_{\mathrm{max}}-M_{\mathrm{ecl}}$)
relation, which is also observationally established \citep[see i.~e.][]{2006MNRAS.365.1333W,2013MNRAS.434...84W}.
To describe the galaxy-wide mass distribution of
stars, one has to perform the integral of all clusters, groups and
associations that are forming within a galaxy.  This leads to the
concept of the IGIMF, further developed and described in Sec.~\ref{sec:imf}. 
Since each cluster has a different upper mass $m_{\mathrm{max}}$, the IGIMF
turns out to be different in shape compared to the IMF within each
cluster.  In particular, its slope in the range of massive stars is
steeper than the star cluster IMF one \citep{2005ApJ...625..754W}, but see \citet{2013arXiv1309.6634W} for top-heavy IMFs in starbursts.

Often it is assumed that the IMF can be interpreted to be a
probability density distribution function \citep[i.~e.][]{2001MNRAS.322..231K,
2002Sci...295...82K,2006ApJ...648..572E,2011ApJ...741L..26F}. 
A star cluster is then simply
assumed to be an ensemble of stars, characterized by its number of
stars, $N$, which are randomly drawn from an underlying universal IMF,
implying that the star formation process within a star cluster is a
purely stochastic one. However, \citet{2013MNRAS.434...84W} have shown recently that a fully
stochastic sampling does not reproduce well the observed $m_{\mathrm{max}}-M_{\mathrm{ecl}}$ 
relation.  A different kind of sampling \citep[optimal sampling,][]{2013pss5.book..115K} that populates a star cluster of mass $M_{\mathrm{ecl}}$ with the
optimal number of stars starting from the most massive stars $m_{\mathrm{max}}$, 
reproduces instead very well the observations 
\citep{2006MNRAS.365.1333W,2013MNRAS.434...84W}. This sampling method is based on
the notion that the physics of star-formation is self-regulated in a
resource-limited environment.

We present chemo-dynamical simulations of the evolution of DM-free TDGs in an external tidal field and hot halo gas to investigate the chemical and dynamical feedback of the first star formation episodes. Our method allows us to perform high resolution simulations while accurately tracing the chemical evolution of the TDG. We study the stability of TDGs as dwarf galaxy-sized but DM-free objects during their early formation epoch. We probe the impact of different IMF descriptions on the survivability of TDGs by assuming either fully-populated or truncated IMFs, in line with the IGIMF theory, as the most extreme cases of possible massive-star numbers and their stellar energy feedback. In Sec.~\ref{sec:simulation} we describe the additional modules that have been developed to extend the Flash code to radiative cooling, self-regulated star formation, stellar feedback and an external tidal field. The full setup is used to simulate the early evolution of a TDG and results with data analysis is presented in Sec.~\ref{sec:results} with a discussion and conclusions in Sec.~\ref{sec:discussion}. 

\section{Simulations} \label{sec:simulation}

\subsection{The numerical code}

The simulation code used in this work is based on the adaptive-mesh refinement code Flash \citep{2000ApJS..131..273F}, version 3.3. 
The gas hydrodynamics in Flash can be treated with different operator splitting methods. Beside the directionally split piecewise-parabolic method 
(PPM) also directionally unsplit solvers are available in Flash3.3. We use the provided MUSCL (Monotone Upstream-centered Schemes for Conservation Laws)-Hancock type formulation \citep{1979JCoPh..32..101V}, which is second-order accurate in both time and space. The provided unsplit hydro solver has been reduced significantly in its memory requirements by N. Mitchell. Star clusters are represented by active particles and are advanced by a variable-timestep leapfrog integration (see Flash User guide at flash.uchicago.edu for more information). For phase transitions from the ISM to a star cluster, in the case of star formation, and vice versa for stellar feedback, properties such as energy, mass and chemical abundances are mapped between grid and particles via Cloud-in-Cell mapping. Therefore the weighting of nearby grid points is proportional to the volume of the particle ``cloud" on each grid cell.  An additional module allows for the creation of star cluster particles during the simulation and the influence of the tidal field.
 The simulation is advanced on the minimum time-step for all cells, determined by the Courant-Friedrichs-Lewy (CFL) condition \citep{1967IBMJ...11..215C}. 
For the simulations presented here, the CFL constant is set to 0.1 and time-steps vary between $\Delta t=[10^3 \dots 3 \times 10^4]\,\mathrm{yr}$.

\begin{figure}
\begin{center}
	\includegraphics[width = \linewidth]{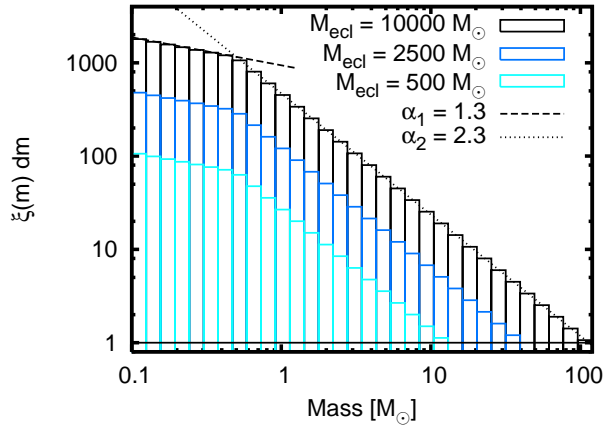}
\caption{\citet{2001MNRAS.322..231K}  IMF with 32 mass bins renormalized for embedded cluster masses of $10^4\,\Msunrm$, $2500\, \Msunrm$, and $500\,\Msunrm$ (from top to bottom row of boxes). $\alpha_1$ and $\alpha_2$ indicate the slope for Eq.~\ref{eq:imf1}.}
\label{fig:imf}
\end{center}
\end{figure}

\subsection{Chemical species and radiative cooling}

The chemical evolution of TDGs is important for finding constraints to classify observed dwarf galaxies as classical dwarfs or old TDGs. 
The mass fractions of 12 different elements: H, He, C, N, O, Ne, Mg, Si, S, Ca, Fe and X (sum of all remaining species) are traced during the simulation. 
All species are advected with the gas flow and are allowed to mix during this process. For every grid cell the abundances are known and radiative 
cooling can be accurately calculated. For gas temperatures above $10^4\, {\mathrm K}$, \citet{1989AA...215..147B} calculated the cooling rates in dependence of the contributions to individual elements.
Together with the cooling rates by \cite{1972ARAA..10..375D} for temperatures below $10^4\, {\mathrm K}$, the radiative losses for every time-step are solved iteratively 
by an implicit Newton-Raphson scheme. 
The electron density and the ionization fraction $f_i = n_e / n_H$ are derived from the ionic fractions at ionization equilibrium by 
\citet{1992ApJ...398..394A} for iron and \citet{1985AAS...60..425A}  for all other elements. 

\subsection{Star formation}
\subsubsection{Star formation rate} \label{sec:sfr}

Stellar particles representing the stellar populations in the simulation are formed in case of a convergent flow. The star formation is fully self-regulated and follows the analytical description
by \citet{1995AA...296...99K} with the stellar birth function $\Psi$ 

\begin{equation}
	\Psi(g, T) = C_n g^n f(T) \, ,
\end{equation}

\noindent with 

\begin{equation}
	f(T) = e^{-T/T_s} \, ,
\end{equation}

\noindent where $g$ and $T$ are gas density and temperature. As derived by \citet{1995AA...296...99K} for collisionally excited radiative cooling, $n = 2$ and we set $C_2 = 2.575 \times 10^8$ (in cgs units) and $T_s = 1000\,{\mathrm K}$ according to the prescription of \citet{1969MNRAS.145..405L}.

During the simulation, whenever the star formation criteria are fulfilled and there is no other stellar particle within a radius $R_{\mathrm {mc}}$, a scalable parameter representing a typical size of a molecular cloud, a new stellar particle is created. For the simulations presented here, we use $R_{\mathrm{mc}} = 220\,{\mathrm{pc}}$ which is equal to 3 times the grid spacing. $R_{\mathrm{mc}}$ is limited by the number of guard cells communicated between blocks on different processors. Flash exchanges 4 guard cells in every dimension and the particles within this region. An $R_{\mathrm{mc}}$ of a maximum of 3 times the grid spacing on the highest refinement level ensures that the particle mapping is correct in the block boundaries.

\begin{figure}
\begin{center}
	\includegraphics[width = \linewidth]{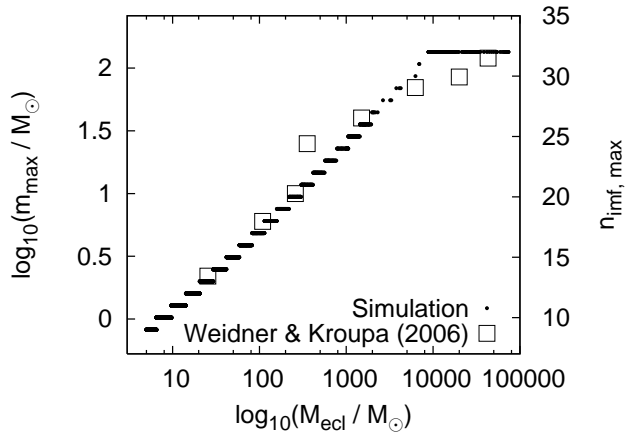}
\caption{Enclosed cluster mass to maximal star mass relation for all stellar particles (black dots) after a simulation time of 500 Myr covering different SF epochs. On the left y-axis the corresponding star mass of the last populated IMF bin is indicated. The right y-axis shows $n_{\mathrm{imf,max}}$, the maximum IMF mass bin that is still populated by at least one star. The steps visible in the data points resemble the positions when the next IMF bin is filled.  Observed maximal star masses and embedded cluster masses for the star clusters listed in Table A1 of \citet{2006MNRAS.365.1333W} are indicated as squares.}
\label{fig:mecl_mmax}
\end{center}
\end{figure}

The material transfer from the gas phase to the stellar particle is given by the analytic description of the stellar birth function. The transfer back to the gas phase by stellar feedback is dependent on mass, age, metallicity and composition of the underlying stellar population as explained in Sec.~\ref{sec:feedback}. Each stellar particle therefore is represented by two different masses: \\
\noindent
1.) The mass of the total stellar population or the embedded cluster mass $M_{\mathrm{ecl}}$ is the mass that the particle has at the time when it is closed for further SF. $M_{\mathrm{ecl}}$ stays constant for the rest of the simulation, scales the IMF and determines the maximal star mass in the case of the truncated IMF (see Fig.~\ref{fig:imf}).

\noindent
2.) As the star cluster ages, stars die and material is returned to the ISM. This mass is therefore time-dependent and used to calculate the gravitational potential and motion of the stellar particles.
The initial velocity is mapped from the gas velocity defined on the grid onto the star particle but afterwards each particle moves independently in the tidal field. 
During a cluster formation time of $\tau_{\mathrm {cl}}$ \citep[here: $\tau_{\mathrm {cl}} = 10\, \mathrm{Myr}$, see e.g.][]{2004MNRAS.350.1503W} grid cells within $R_{\mathrm {mc}}$ do not create new stellar particles but the mass of the already existing particle is increased. 

In regions with a high local SFR, young massive stars heat up their environments shortly after they are born by stellar winds and can regulate further SF. 
We allow for this type of self-regulation in reducing the delay between the cluster formation and the stellar feedback processes for high SFRs. This is numerically realized by starting the stellar feedback (``closing" the particle for further SF) within $\tau_{\mathrm{cl}}$ if $M_{\mathrm{ecl}}$ is high enough so that all mass bins of the IMF are populated with at least one star. For the chosen binning the highest mass bin within the IMF contains at least one star, if $M_{\mathrm{ecl}} > 8840\,\Msunrm$.

Therefore, either after the lifetime of the particle has reached $\tau_{\mathrm {cl}}$ or after the IMF is completely filled, the stellar particle is closed, no further star formation on this particle is allowed and the feedback processes start. When the star formation criteria is still fulfilled in nearby grid cells, a new particle is created. In this case the star cluster consists of more than one stellar particle with the advantage that the newly born OB stars already influence the SF around them. This only happens during the episode of the central starburst, as seen in Fig.~\ref{fig:xiecl}, where the resulting embedded cluster mass function (CMF) for the simulation run with truncated IMFs is shown. For spatially distributed SF, as for example in the first 100 Myr of the simulation, the CMF is populated only up to cluster masses of a few thousand $\Msunrm$. During the episode with the most concentrated SF, between t = 150 and 200 Myr, also star clusters in the mass range between $10^4$ and $10^5\,\Msunrm$ form. This is possible if the SFR is high enough to build massive clusters in a single timestep. 

Based on empirical derivations of the mass function of young clusters with masses between $10^2$ and $10^5\,\Msunrm$, the CMF is assumed to follow a continuous power law of $\mathrm{d}N / \mathrm{d}M_{\mathrm{ecl}} \propto M_{\mathrm{ecl}}^{-\beta}$ with an index $\beta$ close to $2$ with a possible exponential truncation at higher masses \citep[for details see i.~e.][]{2009MNRAS.394.2113G}. \citet{2003AJ....126.1836H} concluded that  $\beta \approx 2.4$ for the Large and Small Magellanic Clouds (LMC, SMC) while \citet{2006MNRAS.366..295D} find indices of $\beta = 1.8 \pm 0.1$ (LMC) and $\beta = 2.00 \pm 0.15$ (SMC) as well as hints for a time-dependent CMF. In our simulations the CMF is not enforced but forms self-consistently because of the individual local gas properties. In Fig.~\ref{fig:xiecl} the resulting CMF for different episodes during the simulation are indicated. We chose episodes, where depending on the local SFR density, either massive clusters with $M_{\mathrm{ecl}} > 10^4\,\Msunrm$ or only low-mass clusters form.

\subsubsection{Initial mass function} \label{sec:imf}

For the simulations presented here we do not aim at resolving individual stars but assume that every stellar particle contains a star cluster of various sizes. The mass of a stellar particle depends on the star formation rate in the volume where it was formed. For a detailed study of the feedback processes of the formed stars, it is important to know not only the total mass of stellar material but also the distribution of stellar masses. At cluster formation we assume an IMF for the underlying population of individual stars with 

\begin{equation} \label {eq:imf1}
	\xi(m) \propto m^{-\alpha_i} \, .
\end{equation}

\noindent We use the multiple-part power-law IMF by \citet{2001MNRAS.322..231K} with:

\begin{eqnarray*}
	\alpha_1 &= & 1.3 \qquad 0.1 \le  m/\Msunrm  < 0.5  \, , \\
	\alpha_2 &= & 2.3 \qquad 0.5 \le    m/\Msunrm \, .
\end{eqnarray*}

\noindent
The IMF is logarithmically divided into 32 bins in the mass range of 0.1 to 100 $\Msunrm$ (see Fig.~\ref{fig:imf}).
For regions with low or moderate SFR the stellar mass, that is built up during the cluster formation time, is not enough to fill the IMF such that there is at least 1 star in the highest mass bin. Numerically the IMF can easily be scaled down to low masses, however leading to fractions of individual high-mass stars being formed, but fractions of massive stars in full stellar populations are unphysical and also do not obey the observed relation between embedded cluster mass, $M_{{\mathrm {ecl}}}$, and the maximal star mass $m_{{\mathrm {max}}}$ in the cluster \citep[see][and references therein]{2006MNRAS.365.1333W,2013MNRAS.434...84W}. 

In addition to the standard approach of a filled IMF in all stellar populations, we performed simulations of TDGs where the IMF is truncated.
Depending on the stellar mass that is accumulated during $\tau_{\mathrm {cl}}$, the IMF is filled only up to the highest mass bin where the number of stars is still $\ge 1$ (see Fig.~\ref{fig:imf} for different $M_{{\mathrm {ecl}}}$). We represent the full IMF with $n_{{\mathrm {bin}}} = 32$ mass bins, and the position of the truncation is given by $n_{{\mathrm {imf,max}}}$, the index of the highest populated mass bin. This way, we reproduce the observed relation between the embedded cluster mass and the maximal star mass. 
The resulting $M_{{\mathrm {ecl}}}$ - $m_{{\mathrm {max}}}$ data for all stellar particles after a simulation time of 500 Myr is shown in Fig.~\ref{fig:mecl_mmax}. The populated region matches the analytical, semi-analytical and numerical trends from previous studies as summarized in Fig. 1 of \citet{2006MNRAS.365.1333W}. 

Observed galaxies as well as the simulated TDGs in this work consist of regions with different local star formation rates and therefore different cluster masses. Subsequently, the integrated IMF is composed of star cluster IMFs with truncations at different star masses. The integrated, galactic IMF \citep[IGIMF][]{2003ApJ...598.1076K} is described by

{\small
\begin{equation}  \label{eq:igimf}
	\xi_{\mathrm{IGIMF}}(m) = \int_{M_{\mathrm{ecl,min}}}^{M_{\mathrm{ecl,max}}} \xi(m<m_{\mathrm{max}}(M_{\mathrm{ecl}})) \xi_{\mathrm{ecl}}(M_{\mathrm{ecl}}) \mbox{d} M_{\mathrm{ecl}} \, ,
\end{equation} }
 
\noindent where the embedded clusters are described by their mass function $\xi_{\mathrm{ecl}}(M_{\mathrm{ecl}})$ and the minimum and maximum cluster masses, $M_{\mathrm{ecl,min}}$, and $M_{\mathrm{ecl,max}}$, respectively. The stellar IMF in each of these clusters is only filled up to a maximal star mass $m_{\mathrm{max}}$. At episodes with low SFRs the maximum mass of embedded clusters, $M_{\mathrm{ecl,max}}$, is smaller, leading to more truncations in the stellar IMF and a steeper $\xi_{\mathrm{IGIMF}}(m)$ \citep{2005ApJ...625..754W}.

\begin{figure}
\begin{center}
	\includegraphics[width = \linewidth]{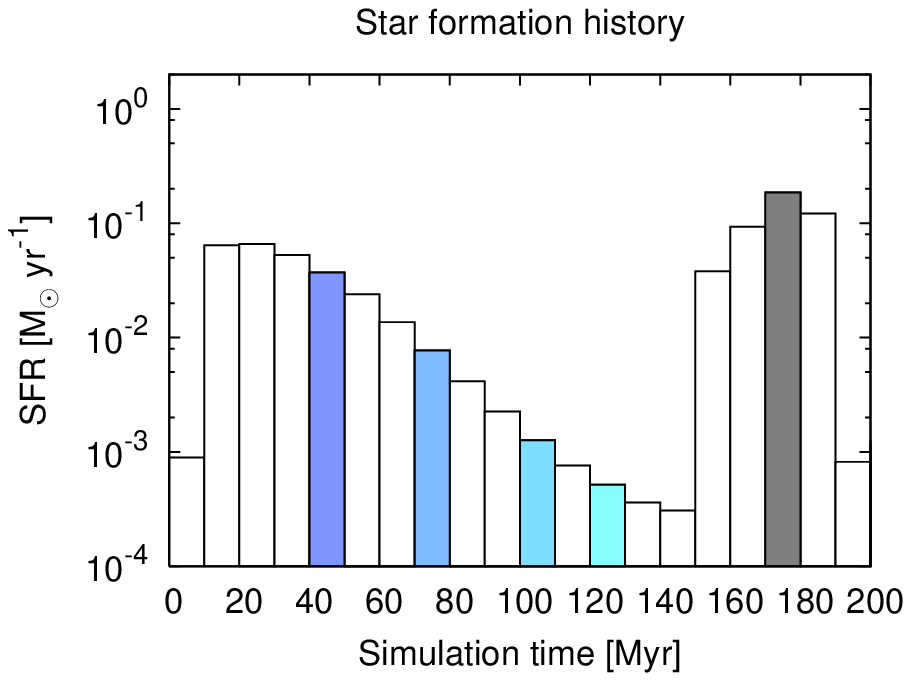}
	\includegraphics[width = \linewidth]{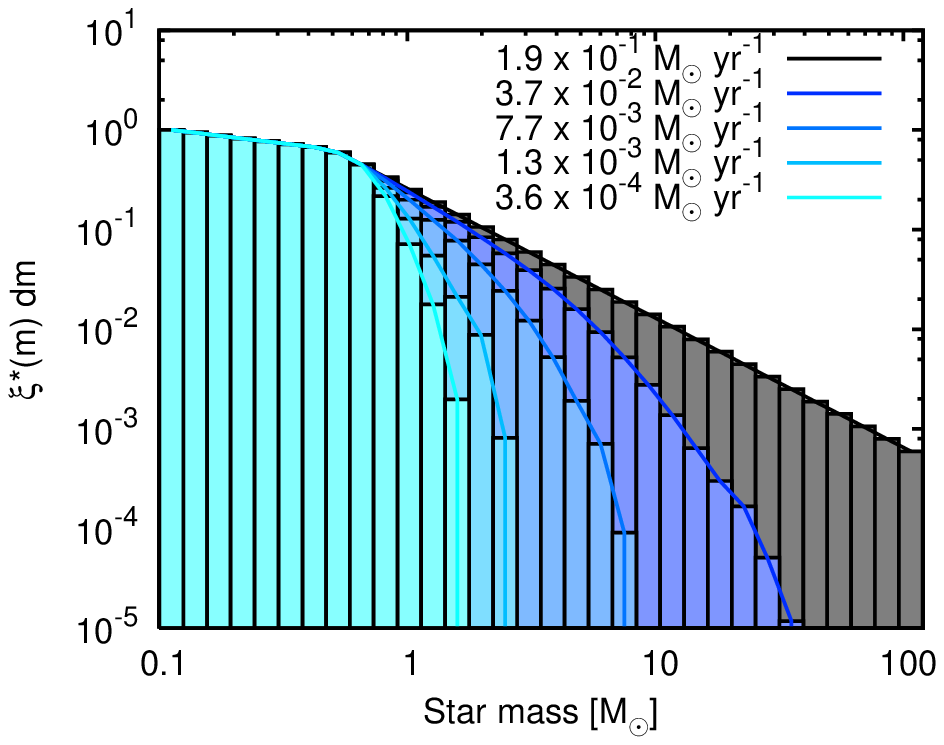}
\caption{Top panel: Star formation history of the first 200 Myr of the simulation with the truncated IMF. For five selected time bins with different star formation rates, the IGIMF is calculated and shown in the lower panel. Bottom panel: IGIMF for clusters with $M_{\mathrm ecl} > M_{\mathrm ecl,min} = 5\, \Msunrm$  at times with different total star formation rates. The number of stars, $\xi^{*}(m)\, \mathrm{d}m$, is scaled to the number of stars in the first mass bin. The colors in this figure represent the chosen stages in the star formation history in the top panel. }
\label{fig:igimf}
\end{center}
\end{figure}

\begin{figure}
\begin{center}
	\includegraphics[width = \linewidth]{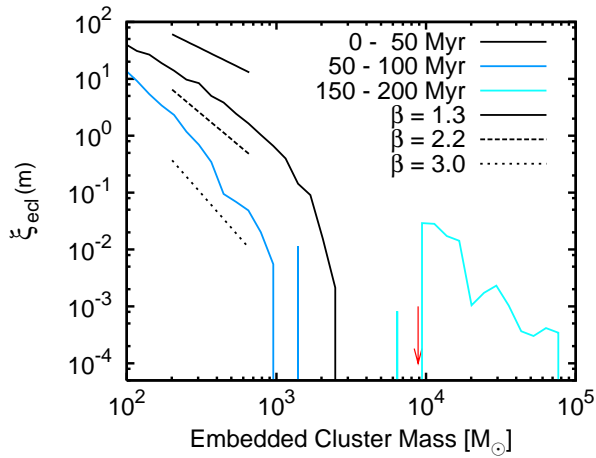}
\caption{  The embedded cluster mass function (CMF) for the truncated IMF run. The mass functions for all clusters formed between the simulation times of [$0 \dots 50$], [$50 \dots 100$], and [$150 \dots 200$] Myr are plotted. Each episode represents different total SFRs.  The power law slopes for the CMF of $\beta = 1.3,\, 2.2, \, \mathrm{and}\, 3.0$ are indicated for demonstration (see Sec.~\ref{sec:sfr}). The arrow indicates the minimum cluster mass ($8840 \,\Msunrm$) where at least one star populates the highest star mass bin. }
\label{fig:xiecl}
\end{center}
\end{figure}

In the simulations presented here, the only relation that is pre-set in the code is the form of the stellar IMF (eq.~\ref{eq:imf1}), while the rest is dependent on the local behavior of the gas. In high-density and low temperature regions massive clusters with filled stellar IMFs are created, while at the same time in a region with a low SFR only low-mass clusters with truncated IMFs are produced. Throughout the first 200 Myr the simulated TDG shows a vivid star formation history with times of low SFRs of the order of $10^{-4}\,\Msunrm\,\mathrm{yr}^{-1}$ to small starburst episodes with $\mathrm{SFR} = 0.2\, \Msunrm\, \mathrm{yr}^{-1}$. In Fig.~\ref{fig:igimf} the IGIMFs are presented for different SF episodes during the first 200 Myr.

\subsubsection{Truncated IMF - IGIMF: }
Although similar in the final results, the IGIMF \citep{2003ApJ...598.1076K, 2005ApJ...625..754W, 2013pss5.book..115K}
and the truncated IMF we introduce in
this paper differ in the main underlying assumptions.  The IGIMF is
based on the fundamental assumptions that ($i$) most stars form in
small stellar groups, i.e. in embedded clusters.  Within each cluster,
the stars are distributed according to a universal multi-slope power
law IMF (the so-called Kroupa IMF).  The maximum stellar mass in an
individual cluster is a function of the cluster mass; ($ii$) the
stellar cluster masses are distributed according to a single-slope power law
and ($iii$) the maximum possible mass of a star cluster increases with
the galactic SFR.  The combination of these three assumptions lead to
the IGIMF formulation as expressed in Eq.~\ref{eq:igimf}.  The main
implication is that dwarf galaxies, characterized by low SFRs, present
a steeper (top-light) IMF than larger galaxies.  In this paper, we
keep the first assumption, namely we assume an universal IMF within
each star cluster and a upper mass depending on the cluster mass.  We
do not impose a specific form for the distribution function of stellar
clusters, neither do we assume a dependence of the maximum cluster mass
with the SFR.  We only need to assume a maximum time-scale for the
formation of star clusters, as explained in detail in Sect. \ref{sec:sfr}.  As
we have seen in Fig.\ref{fig:igimf} the results of these less restrictive
assumptions are still qualitatively similar to the IGIMF results.  The
only difference is that we have a clear dependence of the IMF on the
local SFR rather than on the overall SFR (see Sec.~\ref{sec:imfdepsfr}). 
In the Appendix~\ref{Sec:app} we show, by means of simple analytical tests, the effect
of a truncated IMF. In particular, we show the differences in the
produced number of massive stars and yields of heavy elements between
a truncated and a filled IMF.

Although we do not imply the full IGIMF theory in our simulations, we use the term IGIMF
throughout the paper, whenever the individual truncated IMFs are integrated over the
galaxy.

\subsubsection{Stochastic SF: }
Another way to describe the SF is to distribute the total stellar mass of the embedded star cluster into
the IMF bins stochastically. For a large number of star clusters the integrated stochastic IMF resembles 
a fully-populated IMF, while locally the feedback processes allow for a larger spread. 
In addition, the problematic assumption that there are very small fractions of 
stars contributing to the feedback within individual star clusters is not necessary. 
We compare the stochastic IMF with the truncated and the fully-populated IMF in Appendix \ref{app:stochastic}
in a toy model. Stochastically filled IMFs for $10^4$ test clusters with a mass of $500\,\Msunrm$ each, are 
calculated to derive the mean values as well as the standard deviation for the stellar feedback. The 
average values for the number of SNeII events as well as for the total stellar yields represent the values for 
the filled IMFs, as expected. In spite of the large standard deviation, the feedback of the truncated IMF lies 
outside of $1\,\sigma$ compared to the feedback distributions of the stochastic IMFs. As only the global distributions of the filled and the truncated IMF are significantly different, we do not 
perform a full simulation with stochastic IMF filling.

\subsection{Stellar feedback} \label{sec:feedback}
\subsubsection{Energetic feedback}

As soon as a particle is closed for further SF, either after the IMF is filled or after $\tau_{\mathrm{cl}}$, the feedback processes start. The lifetime for stars ($\tau_{\star}(m,Z)$) 
with the average mass of each IMF mass bin is given by the metallicity dependent stellar lifetimes of 
\citet{1998AA...334..505P} and we assume a constant wind feedback in thermal energy $e_{\mathrm {th}}$ for 
high-mass stars ($M \ge 8\, \Msunrm$) during their lifetime as described by \citet{1992AA...265..465T}:

\begin{equation}
	\left . \frac{\partial e_{{\mathrm {th}}}}{\partial t} \right |_{{\mathrm {OB}}} = \frac{1}{2} \dot{m} v_{\infty}^2 + \eta_{{\mathrm {Ly}}}L_{{\mathrm {Ly}}}(m)  \, ,
\end{equation}

\noindent where $\eta_{{\mathrm {Ly}}}$ is the fraction of the mean photon energy of 17~eV that is converted into kinetic energy of the surrounding gas \citep[$\eta_{{\mathrm {Ly}}}=10^{-3}$,][]{1992AA...265..465T}, and a metal-dependent mass loss rate by stellar winds of 

\begin{equation}
	\dot{m} = -10^{-15} \left ( \frac{Z}{\mathrm{Z}_{\odot}} \right )^{0.5} \left ( \frac{L}{\Lsunrm} \right )^{1.6}\, \Msunrm \, \mathrm{yr}^{-1} \, ,
\end{equation}

\noindent and a final wind velocity of

\begin{equation}
	v_{\infty} = 3 \times 10^3 \left ( \frac{m}{\Msunrm} \right ) ^{0.15} \left ( \frac{Z}{\mathrm{Z}_{\odot}}\right )^{0.08}\, {\mathrm {km}}\, {\mathrm s}^{-1} \, .
\end{equation}

\noindent The ionizing radiation of a star with mass $m$ is given by 

\begin{equation}
	L_{{\mathrm {Ly}}}(m) = 10^{40} \left( \frac{m}{\Msunrm} \right)^6\, \mathrm{photons\,s}^{-1} \mathrm { star}^{-1} 
\end{equation}

\noindent \citep{1987MitAG..70..141H}. With this approach the radiative energy from massive stars is included in the sub-grid physics as thermal feedback on the gas phase. For the chosen resolution with a grid size of $(76\,\mathrm{pc})^3$, the mass in stars is only a fraction of the gas mass in each grid cell. Therefore we do not take into account the contribution of radiation pressure from individual stars, as the thermal pressure of the ISM dominates over radiation pressure on this scale.  

\begin{figure}
\begin{center}
	\includegraphics[width = \linewidth]{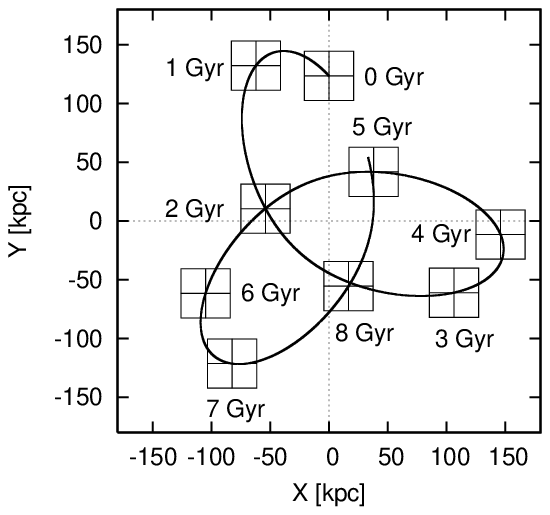}
\caption{The orbit of the simulation box (SB) in the rest-frame barycenter of the merging galaxies (X = 0, Y = 0) is indicated together with positions of the SB at different times. The simulation starts at X = 0, Y = 123.5 kpc, and the SB is moving in the plane of Z=0.}
\label{fig:orbit}
\end{center}
\end{figure}

Each IMF mass bin with an average mass of more than $8\, \Msunrm$ contributes to the wind energy input as long as the stellar 
particle is younger than $\tau_{\star}(m,Z)$. If during the timestep one or more mass bins exceed $\tau_{\star}(m,Z)$, 
all stars in these bins explode as Supernova (SN) type II with immediate energy $E/\mathrm{SNII} = \epsilon\cdot 10^{51}\, {\mathrm {erg}}$, with $\epsilon = 0.05$ \citep[see][]{2001MNRAS.322..800R,2013AA...551A..41R}
and material release (see Sec.~\ref{sec:yields}). 
Intermediate mass stars with $m = [3 \dots 8]\, \Msunrm$ expel their processed material during the AGB phase at the end 
of their stellar evolution. A fraction $f_{\mathrm {bin}} = 0.05$ \citep{1998AA...334..505P} of stars are assumed to be in a close binary system and 
end their evolution in a SNIa explosion with an energy input of ($E/\mathrm{SNIa} = \epsilon\cdot 10^{51}\, \mathrm{erg}$, with $\epsilon = 0.05$).
All stars in intermediate mass bins release the chemical elements produced by stellar nucleosynthesis at the end of their
metal dependent lifetime, analogously to the material ejection of SNII explosions by massive stars (see Sec.~\ref{sec:yields}).

\begin{figure*}
 	\begin {minipage}[b]{0.47\linewidth}
		\includegraphics[width=\linewidth]{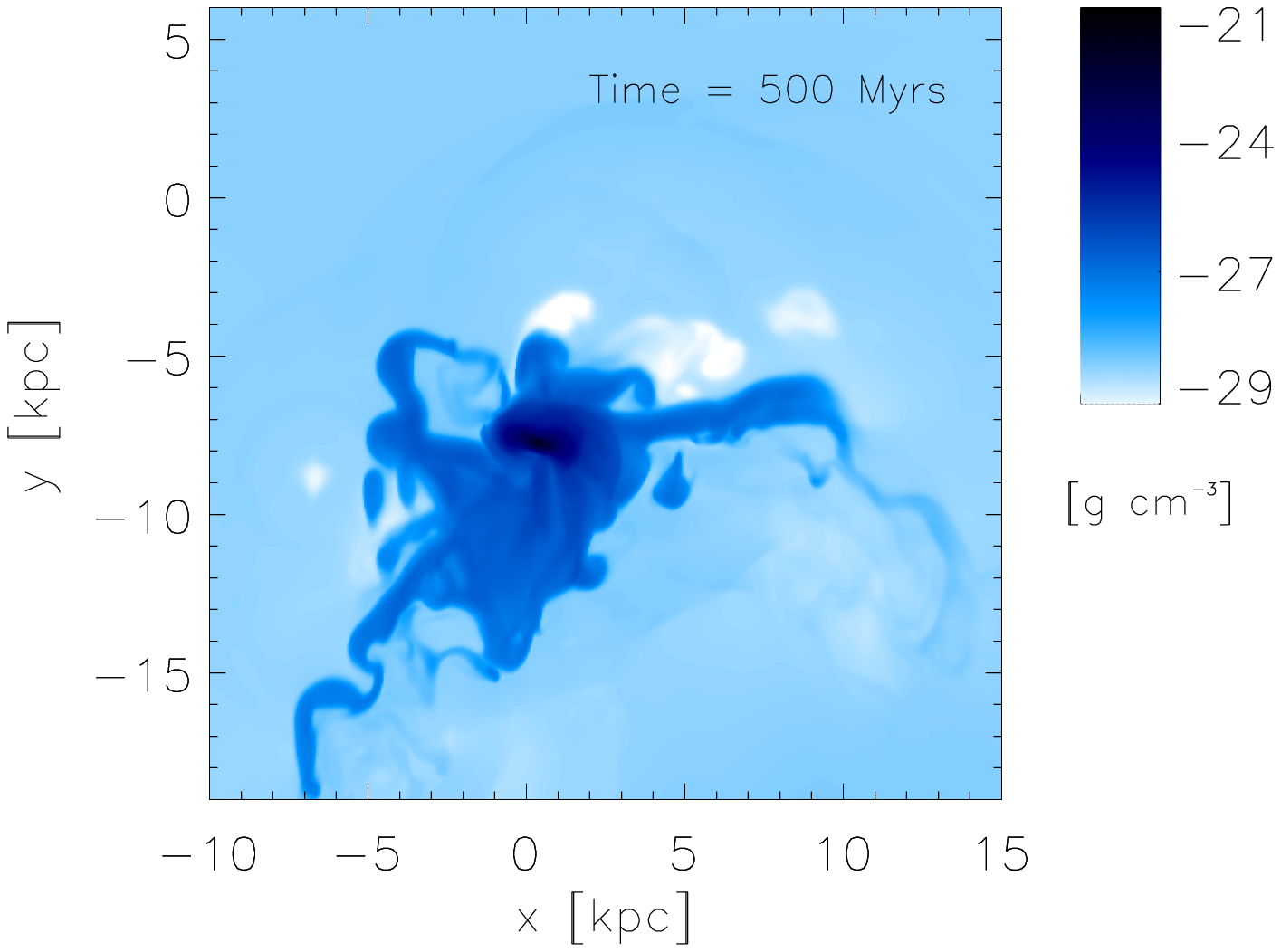}
	\end {minipage}
	 \begin {minipage}[b]{0.47\linewidth}
		\includegraphics[width=\linewidth]{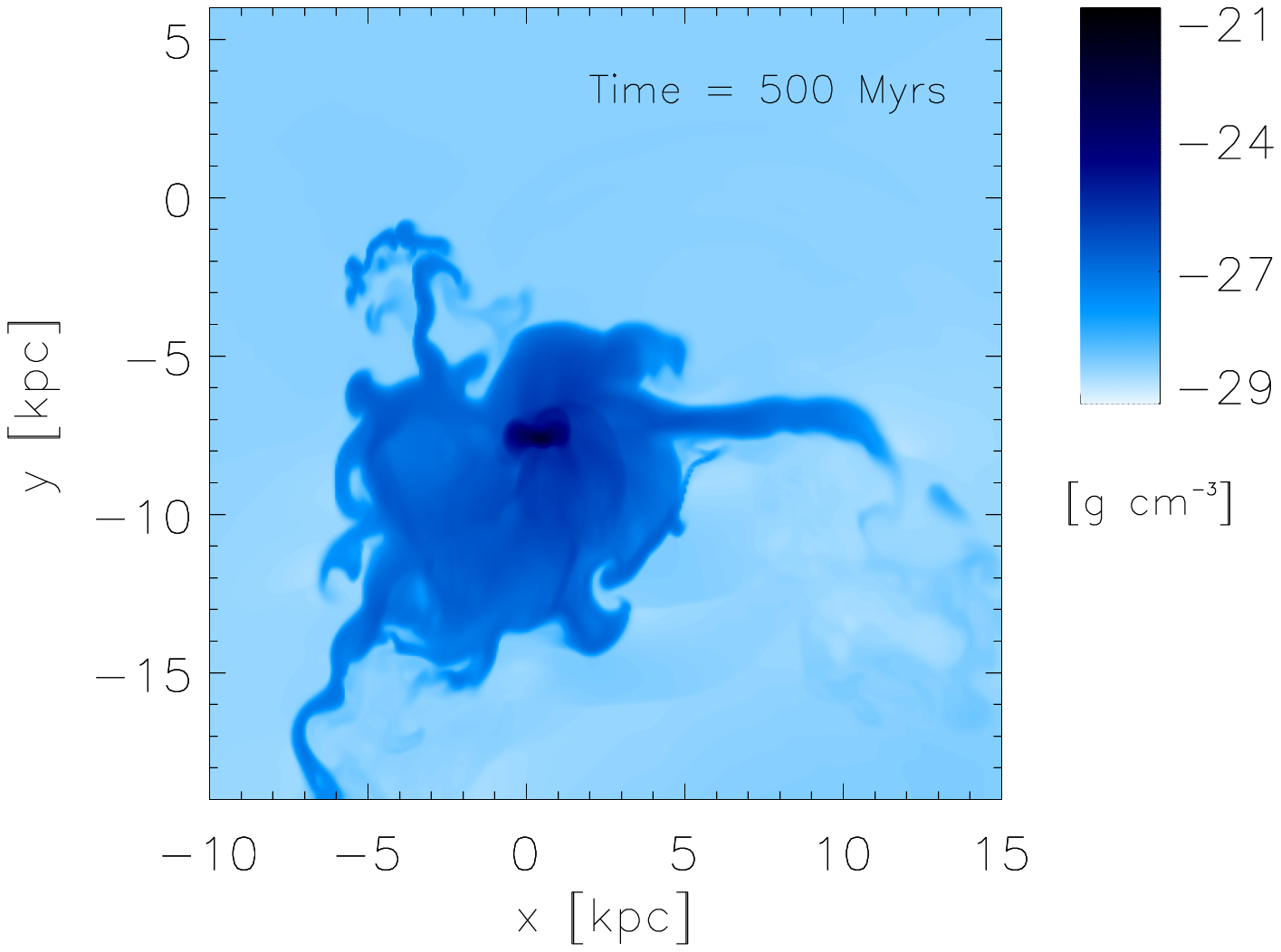}
	\end {minipage}
	
	 \begin {minipage}[b]{0.47\linewidth}
		\includegraphics[width=\linewidth]{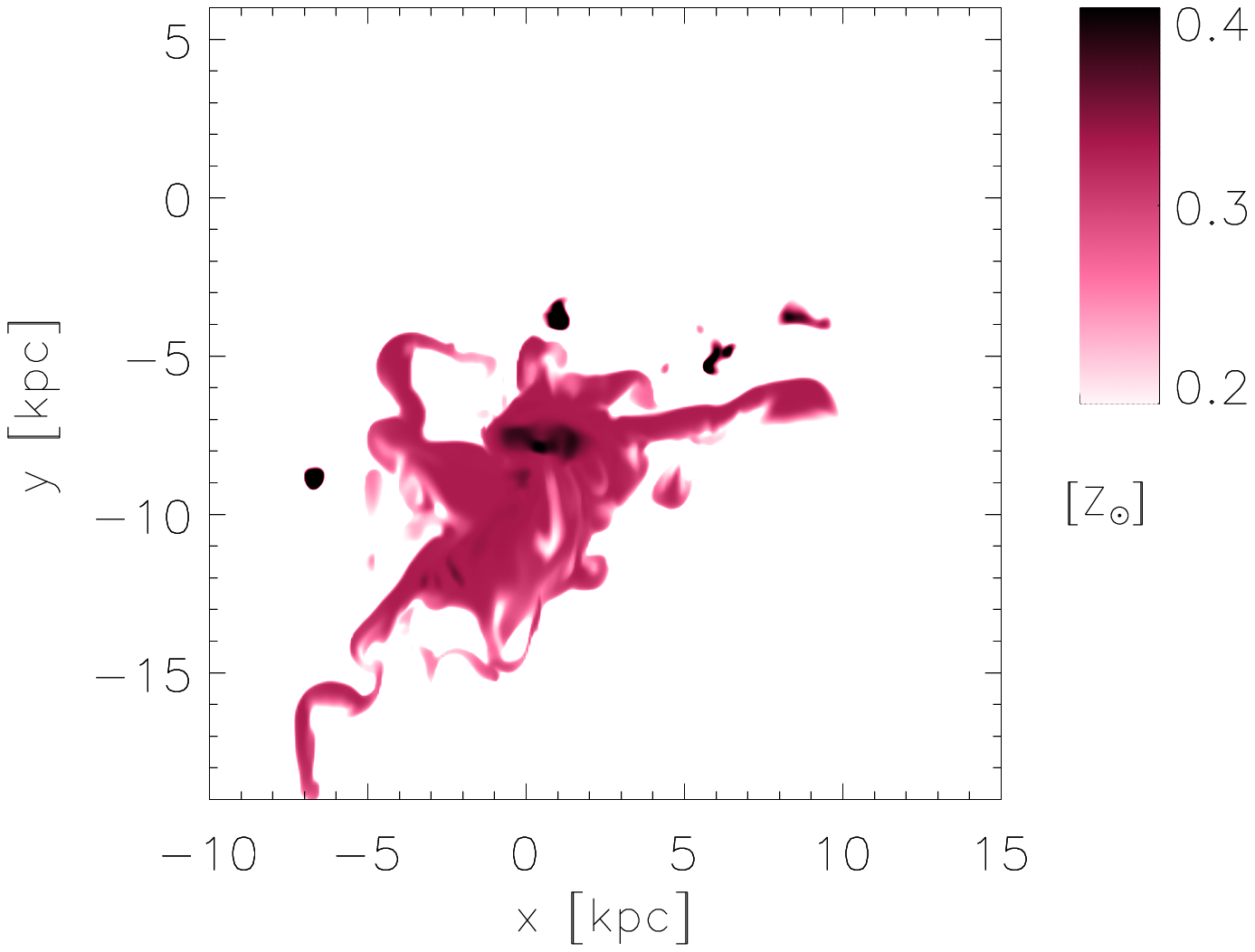}
	\end {minipage}
	 \begin {minipage}[b]{0.47\linewidth}
		\includegraphics[width=\linewidth]{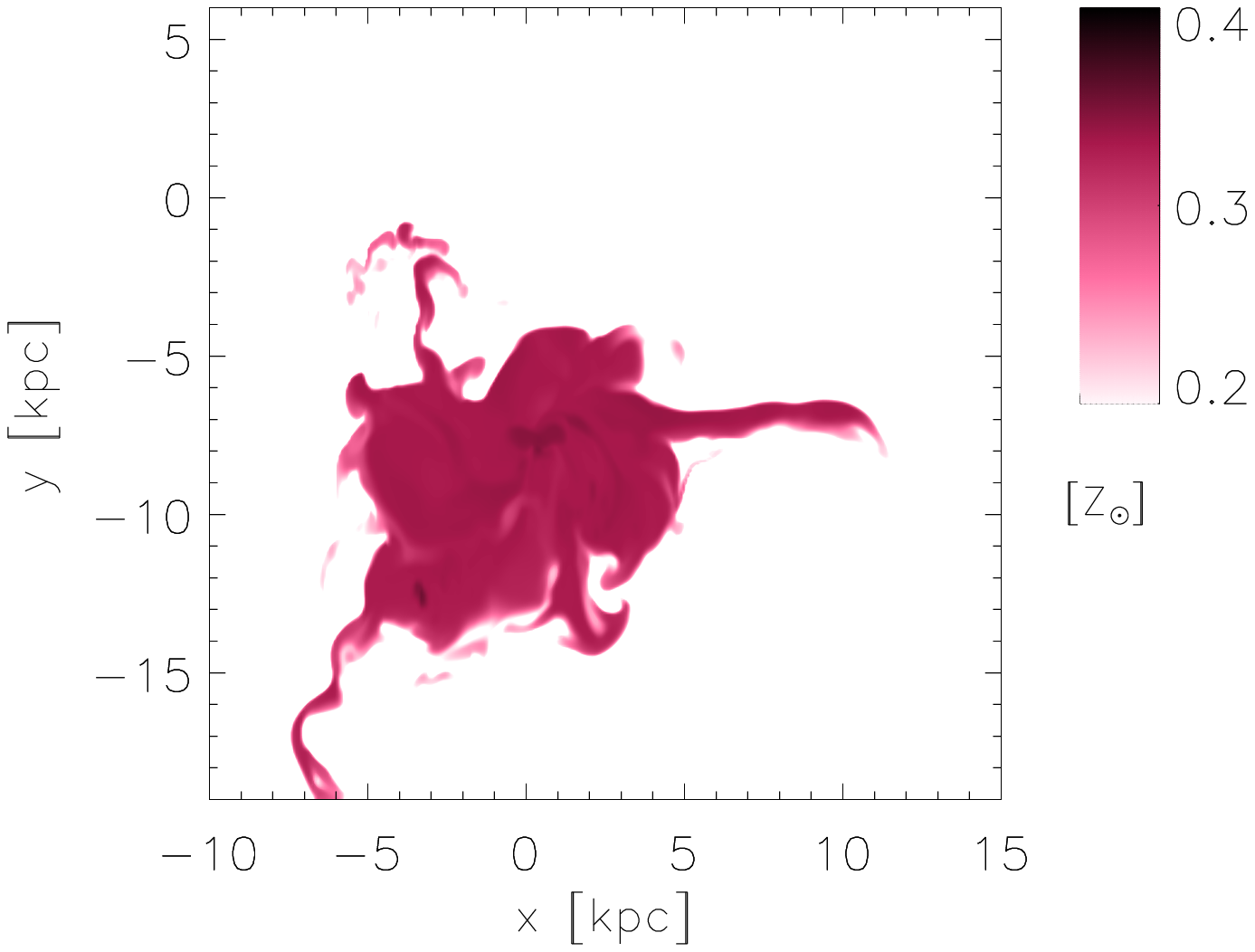}
	\end {minipage}
	\caption{2D slice at z = 0 through the 3D simulation box at t = 500 Myr for the run with the truncated IMF (left) and the fully populated IMF (right). Color coded is the total gas density in g cm$^{-3}$ in the top panels and the gas metallicity in $\mathrm{Z}_{\odot}$ in the bottom panels.}
\label{fig:dens0500}
\end{figure*}

The energy release by SNIa and SNII explosions is considered to be fully thermal, which typically leads to the so-called 
``overcooling problem" because of the unresolved underlying multi-phase structure of the interstellar medium. Averaging over the dense, cold phase and the hot dilute phase in temperature and density leads to cooling rates that are too high for the hot gas. 
While the dense cores of cold gas cool faster, the hot but dilute gas that is heated up and driven by the SN energy has a 
longer cooling timescale. 

 Regulating the radiative cooling after SNe is a common technique to account for this effect. In Lagrangian codes the cooling of a certain number of surrounding gas particles can be switched off for a fixed cooling shut-off time \citep{2006MNRAS.373.1074S, 2011MNRAS.410.1391A} or a timescale proportional to the Sedov solution of the blast wave equation \citep{2010Natur.463..203G}. A similar approach for grid codes is described in \cite{2013MNRAS.429.3068T}, where the cooling rate is set to zero in all grid cells, where the turbulent velocity dispersion is above a given threshold.

In our simulations we avoid strict barriers where radiative cooling is permitted in grid cells under certain
conditions after SNe events. Here, the mass fraction of the released SN material is traced as a mass scalar which is 
advected with the gas, this fraction is considered to be in the hot phase and we only take the radiative cooling of the cold and
warm phase into account. The mass scalar ``snwi" typically spreads over neighboring grid cells within the subsequent 
time-steps and can mix and therefore dilute with the environment of the SN explosion. The radiative cooling for the hot phase
on a longer time-scale is represented by a reduction of the mass scalar ``snwi" according to

\begin{equation}
   \mathrm{snwi}^{(n+1)} =  1.0 - \frac{\Delta t}{\tau_{\mathrm{SN}}} \times \mathrm{snwi}^{(n)} \, ,
\end{equation}

\noindent where $\mathrm{snwi}^{(n+1)}$ and $\mathrm{snwi}^{(n)}$ are the values for the mass scalar at the new and the old timestep,
separated by the time-step $\Delta t$ and $\tau_{\mathrm{SN}} = 3\, \mathrm{Myr}$ represents a typical cooling timescale for the hot phase.

\subsubsection{Mass loss and stellar yields} \label{sec:yields}

The abundance history of 12 chemical species is traced during the simulation. The metallicity of the TDG
is updated by mixing inside the galaxy as well as by stellar feedback from SNIa and SNII explosions and stellar winds during the AGB phase.
We use the metallicity dependent yields by \cite{1996AA...313..545M} for stars with $m = [3 \dots 4]\, \Msunrm$ and 
\cite{1998AA...334..505P} for stars with masses above $6\, \Msunrm$ with a linear interpolation between 4 and 6 $\Msunrm$
stars.

The initial metallicity of the star particle is mapped from the gas on the grid to the stellar metallicity defined on the particle
at the time of the creation of the stellar particle. Therefore even if the gas is significantly enriched during the lifetime of the stellar
population, the stellar metallicity stays independent and can be used to determine the metallicity dependent yields. In turn, the updated 
chemical composition of the ISM after stellar feedback influences the radiative cooling and therefore the dynamics of the ISM.

\subsection{Tidal field} \label{sec:orbit}
Contrary to the isolated evolution of classical dwarf galaxies in low density environments, TDGs form and develop influenced by
a tidal field created by the interacting host galaxies from which the tidal arms are expelled. We focus here on the evolution of
the TDG rather than the interaction process itself as was already done i.e. by \citet{2006AA...456..481B}, \citet{2007MNRAS.375..805W}, 
and \citet{2009ApJ...706...67R}, 
but we do not want to neglect important environmental effects either. Therefore we put the simulation box (SB) in the reference frame of the 
TDG and furthermore on an orbit around the mass center of a host galaxy. The self-gravity within the SB is determined
by the Multigrid Poisson solver of Flash3, that is based on \citet{2008ApJS..176..293R}. In addition the accelerations by the 
attraction of the host galaxies is calculated. During interaction processes the baryonic matter of galaxies as 
well as their dark matter halos are heavily distorted. The external gravitational potential is therefore time-dependent in the formation and 
very early evolution of the TDG. 

We focus in this work on the later evolution when the interacting galaxies are in a late stage of 
their merger event and their gravitational potential has settled. Thus we assume a constant external gravitational potential during
the simulation time. As a typical potential in the outskirts of massive galaxies we chose the DM distribution by \citet{2008ApJ...684.1143X}, who fitted the kinematics of Milky Way halo stars to an NFW profile \citep{1997ApJ...490..493N}:

\begin{equation}
	\Phi_{\mathrm{NFW}} (r) = - \frac{4 \pi G \rho_s r_{\mathrm{vir}}^3}{c^3 r} \, \mathrm{ln} \left(	1 + \frac{cr}{r_{\mathrm{vir}}}	 \right ) \, ,
\end{equation}

\noindent where $\rho_s$ is a characteristic density given by

\begin{equation}
	\rho_s = \frac{\rho_{cr} \Omega_m \delta_{th}}{3} \frac{c^3}{\mbox{ln} (1+c) - c/(1+c)} \, ,
\end{equation}

\noindent $\rho_{cr} = 3H^2 / (8 \pi G)$ the critical density of the universe, $\Omega_m$ the contribution of matter to the 
critical density and $\delta_{th}$ the critical overdensity at virialization. \citet{2008ApJ...684.1143X} used $\Omega_m = 0.3$, 
$\delta_{th} = 340$ and $H_0 = 65\, \mbox{km}\,s^{-1}\,\mbox{Mpc}^{-1}$ and found a good fit for a virial radius $r_{vir} = 275\, \mbox{kpc}$ and a
concentration parameter $c = 6.6$ resulting in a virial mass of $M_{vir} = (4\pi /3) \rho_{cr} \Omega_{m} \delta_{th} r_{vir}^3 = 1.0 \times 10^{12}\, M_{\odot}$.
We use this profile for the external merging galaxies. 

The eccentric orbit of the SB leads to additional accelerations by non-inertial forces, which appear in the rest-frame of the TDG. 
Both the gas, which is defined on the grid, as well as the particles representing the stellar population are influenced by 
the centrifugal and Coriolis force, which are applied on both components at every timestep.

We assume a constant ambient halo gas with a temperature of $T_{h} = 10^6\, \mathrm{K}$, a density of $n_{h} = 4.4 \times 10^{-6}\, \mathrm{H\, cm}^{-3}$ and 
abundances of [x/H] = (-1.75, -1.75, -1.75, -1.75, -1.78, -1.69, -1.75, -1.74, -2.00) for (C, N, O, Ne, Mg, Si, S, Ca, and Fe) respectively \citep[from][]{2012ApJ...753...64I}.
In the rest-frame of the TDG the ambient medium moves through the SB, similar to a wind-tunnel simulation, but the absolute value of 
the wind as well as the direction change according to the motion of the box on its orbit. This way, we also include effects caused by
ram pressure stripping. 

The initial position and velocity of the SB are chosen to approximately represent the orbit of TDGs in the simulations 
by \citet[][see their Fig.~1]{2006AA...456..481B}. In the coordinate frame of the interacting galaxies, the SB starts at a 
distance of $R_{\mathrm{SB}} = 123.5\, \mathrm{kpc}$, an angle of $\theta_{\mathrm{SB}} = 90^{\circ}$, and a 3D velocity of 
$\overrightarrow{v}_{\mathrm{SB}} = (-71.7, 78.2, 0)\, \mathrm{km}\,\mathrm{s}^{-1}$. The position of the simulation box at the next timestep
is determined by integration of the motion of the SB within the potential $\Phi_{\mathrm{NFW}}$. The resulting orbit is shown 
in Fig.~\ref{fig:orbit}. The top panels of Fig.~\ref{fig:dens0500} show a density slice through the orbital plane of the simulation box after 500 Myr.

\begin{figure}
\begin{center}
	\includegraphics[width = \linewidth, bb = 100 300 1800 1700, clip]{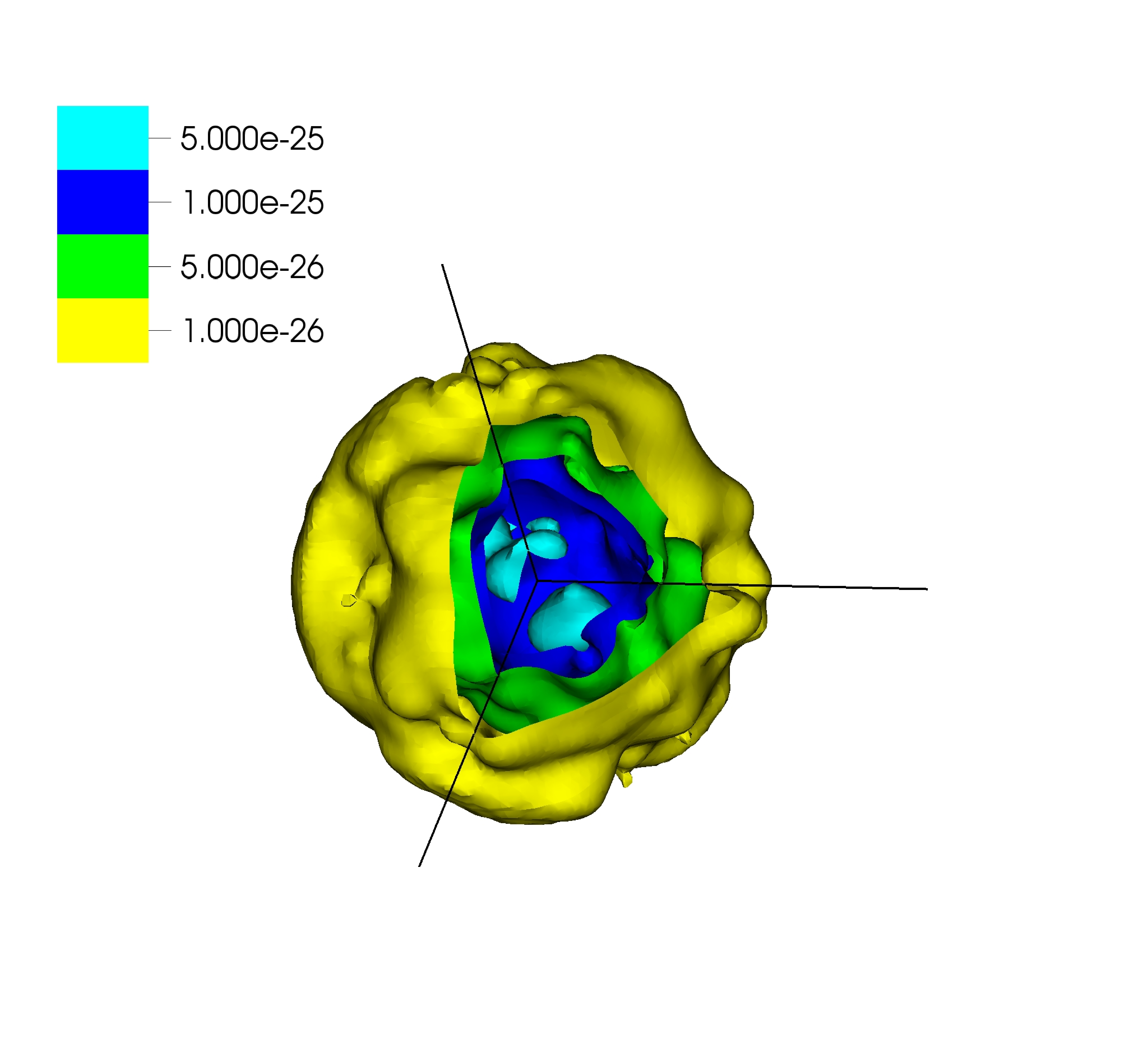}
\caption{Sliced initial 3D density distribution. Isodensity contour are shown for $[10^{-26}, \, 5 \times10^{-26}, \, 10^{-25},\,5\times10^{-25}\, {\mathrm {g\,cm}}^{-3}]$.}
\label{fig:initdens}
\end{center}
\end{figure}

\subsection{Initial setup} \label{sec:init}

The simulation starts at a point where the TDG is kinematically decoupled from the tidal arm and already forms a gravitationally bound 
object. The initial pressure distribution of the gas is in hydrostatic equilibrium as well as in pressure equilibrium with the ambient
hot halo gas. The perfect spherical symmetry is broken in density and temperature by adding a number $n_o =500$ of 
gaussian over-densities with random positions $\vec{r_o}(i) = (x_o(i), y_o(i), z_o(i))$ to the hydrostatic density distribution $\rho(r)$

\begin{equation}
	\rho (r) = \rho(r) + \sum _i^{n_o} A_o(i)  \cdot \rho(r)  \cdot e^{- \frac{|\vec{r}-\vec{r_o}(i)|^2}{2c_o(i)^2}} \, ,
\end{equation}

\noindent with random values for the amplitudes $A_o(i) = [1 \dots 8]$ and $c_o(i)= [100 \dots 600]\,\mbox{pc}$. For a 3D image of the 
initial density distribution see Fig.~\ref{fig:initdens}. Ensuring the hydrostatic equilibrium, the masses of the overdensities are added
to the enclosed mass when the initial radial pressure distribution is constructed. The temperatures and internal energies are calculated from 
the density, pressure and abundance values for each grid cell assuming an ideal gas law with an equation of state in the molar form.
The average atomic mass is calculated individually for every grid cell to account for different abundance ratios.

We assume an initially pressure supported system and therefore add an initial velocity field onto the TDG. The center of each 
overdensity has an initial velocity with an absolute value representing the circular velocity around the center of the TDG at its radius $r$ with a random 
direction in the plane perpendicular to the direction to the center. 

The SB has an extent of 39 kpc and with a maximum refinement level of 6 for blocks with $16^3$ cells, the effective resolution is 76 pc \citep[see][for details on the adaptive mesh refinement of Flash]{2000ApJS..131..273F}. The initial gas mass of the TDG is $M_{\mathrm {Gas,0}} = 1.95 \times 10^8\, \Msunrm$ and no old stellar population is assumed, $M_{\mathrm {Stars,0}} =  0$, in line with observations i.~e.  in Stephan's Quintet where the analyzed spectral energy distribution does not show evidence for a distinct separate old stellar component \citep{2010AJ....140.2124B}. For all simulations presented here, the initial metallicity for the TDG is set to $Z_{\mathrm{TDG}} = 0.3 \, {\mathrm Z}_{\odot}$.

 The initial conditions represent a gas cloud with temperatures below $10^4\,\mathrm{K}$, which leads to an initial starburst episode.
The radiative energy losses between $10^4$ and $10^5\,\mathrm{K}$ are dominated by strong H, He and C lines \citep[see][]{1989AA...215..147B}. Initial temperatures of above $10^4\,K$ result in a rapid
temperature drop. The result is either an even faster collapsing cloud, if the initial equilibrium is not adjusted, or a similar initial starburst with a short time delay caused by the very short cooling timescale.

The initial SFRs in the simulations (up to $6 \times 10^{-2} \, \Msunrm \, \mathrm{yr}^{-1}$) is related to the chosen initial conditions but is in line with observed starbursts in TDGs i.~e. in Arp 105 and NGC 7252, where multi-wavelength observations indicate SFRs of up to $0.1\, \Msunrm \, \mathrm{yr}^{-1}$ \citep[see][for an overview]{2013LNP...861..327D}.

Also a slow gas accretion episode might lead to milder SFRs in TDGs, however in this study, we are interested in the behavior and especially the survivability of TDGs, as low-mass, DM-free galaxies that experience an active SF episode. Therefore, the initial condition are chosen to represent an extreme case in order to investigate whether TDGs get disrupted easily or can survive a starburst and continue with self-regulated SF with a more moderate SFR.

\section{Results} \label{sec:results}

\subsection{IMF dependent SFR}\label{sec:imfdepsfr}

We simulated two initially identical TDGs, where in one simulation run all mass bins over the whole mass range are filled for every star particle, thus allowing fractions of massive stars to be formed. For another simulation the IMF is truncated at the mass bin where the number of stars becomes less than 1. For a detailed description of the different IMF treatments see Appendix~\ref{Sec:app}.
In Fig.~\ref{fig:MLsfh} the time-dependent differences in the star formation history, the total luminosity of the stars, and the masses of stars and gas are summarized.

In the case of the filled IMF massive stars exist in every stellar particle. As the lifetime of stars is inversely proportional to their masses, filled IMFs result in a faster regulation of further SF.
Even if there is only a small fraction of stars with $M_{\star} > 100\,\Msunrm$ in every particle, their quickly released fractional SNII energy feedback increases the 
temperature of the interstellar material and quenches further star formation. In the first 20 Myr of the simulation with truncated IMF, the number of SNeII
is one order of magnitude lower than in the filled IMF case (filled IMF: $2\times 10^3$ SNeII, truncated IMF: $10^2$ SNeII, see Fig.~\ref{fig:SNerate}). In both cases the central part collapses and this leads to an episode of 
relatively high SF after which the SF is regulated again by the feedback of the stars produced in the starburst. 
For the TDG with the truncated IMF this episode starts around t = 150 Myr with SFRs up to $0.2\,\Msunrm\,{\mathrm {yr}}^{-1}$ while the TDG with the 
filled IMF is regulated faster and a similar event at t = 220 Myr only leads to a SFR of around $10^{-3}\,\Msunrm \,{\mathrm {yr}}^{-1}$.

The overall SFR does not fully resemble the individual SF events in the galaxy. In the simulation with truncated IMFs we find comparable SFRs for different times in the evolution of the TDG. As an example we demonstrate in Fig.~\ref{fig:sfnimf} the comparison between the two time spans $t_1= [120 \dots 130]$ Myr and $t_2 = [270 \dots 280]$ Myr. The total mass of stars formed during $t_1$ is 5140 $\Msunrm$ and 4950 $\Msunrm$ during $t_2$, leading to average SFRs of $5.14 \times 10^{-4}\, \Msunrm {\mathrm {yr}}^{-1}$ and $4.95 \times 10^{-4}\, \Msunrm {\mathrm {yr}}^{-1}$, respectively.

\begin{figure}
\begin{center}
	\includegraphics[width=\linewidth]{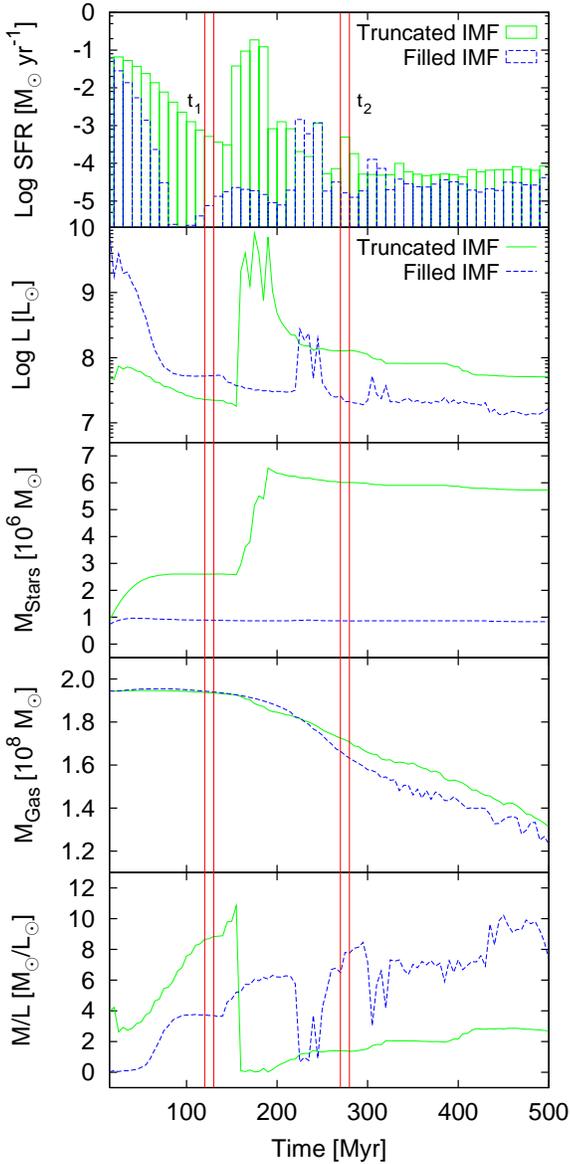}
\caption{Comparison between the simulation with filled (blue dashed lines) and truncated (green solid lines) stellar IMF. The panels show from top to bottom: star formation rate
in bins of 10 Myr , the integrated luminosity of the stellar particles $L$, the total stellar mass $M_{\mathrm {Stars}}$, the bound gas mass $M_{\mathrm {Gas}}$, and the 
resulting mass ($M_{\mathrm {Stars}}+M_{\mathrm {Gas}}$) to light ratio M/L. The time spans $t_1 = [120\dots130]\,\mathrm{Myr}$ and $t_2 = [270\dots280]\,\mathrm{Myr}$ are indicated with vertical solid lines.}
\label{fig:MLsfh}
\end{center}
\end{figure}

If the IGIMF was only dependent on the total SFR, the slope and upper mass limit of the IGIMF is expected to be similar for $t_1$ and $t_2$. In Fig.~\ref{fig:igimf130} the IGIMFs for $t_1$ and $t_2$ are presented, showing very different distributions.

The IGIMF at $t_1$ has an upper mass cutoff in the same order of magnitude as theoretically derived by \citet{2007ApJ...671.1550P} for SFRs between $10^{-4}$ and $10^{-3}\,\Msunrm\,\mathrm{yr}^{-1}$, while the IGIMF of $t_2$ populates the high mass end of the IMF up to stars with masses of $50\,\Msunrm$. Compared to the theoretical IMF truncation by \citet{2007ApJ...671.1550P}, this cutoff mass is expected for much higher SFRs ($\approx 10^{-2}\,\Msunrm\,\mathrm{yr}^{-1}$).

The reason is that during $t_1$ the SF is spread out over a larger volume with typical SFR densities of 
$10^{-5} - 10^{-4}\,\Msunrm\,\mathrm{Myr}^{-1}\,\mathrm{pc}^{-2}$ while in $t_2$ the SF is locally triggered by the collapse of the central core and therefore dominated by high SFR densities (up to $0.1\,\Msunrm\,\mathrm{Myr}^{-1}\,\mathrm{pc}^{-2}$) constrained to the very central part of the TDG (Fig.~\ref{fig:sfnimf}). In this local starburst, massive star clusters with high maximal star masses can be produced, while the lack of SF in the rest of the TDG during $t_2$ leads to the low total SFR of $4.95 \times 10^{-4}\, \Msunrm\,{\mathrm {yr}}^{-1}$.

The differences in the IGIMFs around $t_1$ and $t_2$ are apparent in the complete lack of SNeII around $t_1$ (no SNeII between t= 100 and 160 Myr) while around $t_2$ (between 230 and 320 Myr) the SNII rate is of the order of a few 10 SNeII (10 Myr)$^{-1}$ (Fig.~\ref{fig:SNerate}), even though the total SFRs are comparable. Note that for the simulation run with fully-populated IMFs (blue dashed lines in Fig.~\ref{fig:SNerate}) SNeII are present even for SFRs $< 10^{-5}\,\Msunrm \,{\mathrm{yr}}^{-1}$.

\begin{figure}
\begin{center}
	\includegraphics[width=\linewidth]{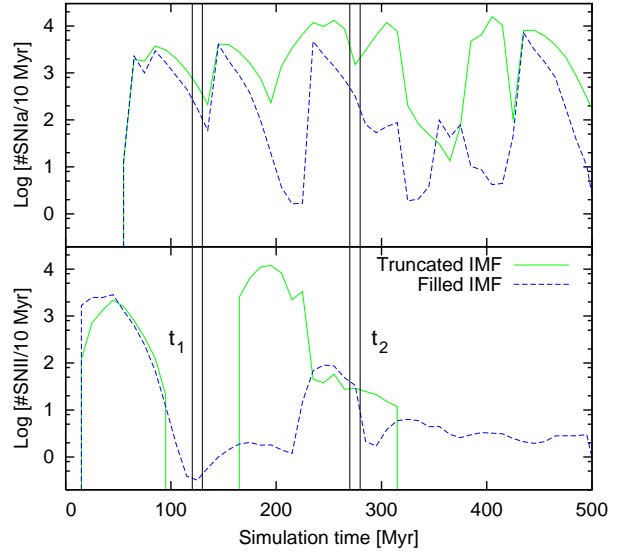}
\caption{Number of SNeIa (top panel) and SNeII (bottom panel) per 10 Myr for the simulated TDG with truncated (green solid line) and filled (blue dashed line) IMFs. Note that for the simulation with the filled IMFs a number of fractional SNII events sums up to the total SNII quantity. The time spans $t_1 = [120\dots130]\,\mathrm{Myr}$ and $t_2 = [270\dots280]\,\mathrm{Myr}$ are indicated with vertical solid lines. }
\label{fig:SNerate}
\end{center}
\end{figure}

We conclude that the condition to truncate the IMF at the mass bin, where the number of stars per bin is smaller than one, reproduces the expected IGIMF for all times where the total SFR is not dominated by individual, local star bursts (see the SFR dependent behavior of the IGIMFs for a selection of time spans without local star bursts in Fig.~\ref{fig:igimf}). In addition, we allow for the formation of massive stars with high $m_{\mathrm {max}}$ in regions with a high local SFR per unit volume even if the integrated galactic SFR at that time is very low, in line with the observed $M_{\mathrm {ecl}}\,-\,m_{\mathrm {max}}$ relation.

The lack or the existence of massive stars influences the gas dynamics and subsequently the further evolution of the galaxy. Massive stars do not only return additional energy to the gas phase by stellar winds, but also the energy injection by SNeII happens on a much shorter timescale (4 - 40 Myr for stars with 100 - 8 $\Msunrm$), compared to the delay between the cluster formation and SNIa ($>> 40$ Myr). For a self-regulated SF, massive stars increase the internal energy of the surrounding ISM faster and consequently further SF can be prevented easier. Vice versa the lack of massive stars postpones the energy feedback by SNe and the local SF can continue for longer.

In order to investigate the stability of the TDG against the stellar feedback processes as well as the tidal field and the ram pressure, we 
calculate the bound gas, $M_{\mathrm {Gas}}$, and stellar mass, $M_{\mathrm {Stars}}$, over the simulation time (see. Fig.~\ref{fig:MLsfh}). 
The bound gas mass $M_{\mathrm {Gas}}$ is determined by a comparison of the corrected potential energy to the sum of the kinetic and internal energy in each cell.
The corrected potential energy reduces the gravitational potential from the simulation by the the external NFW halo to get the 
binding energy of the TDG. The stellar mass, $M_{\mathrm{Stars}}$, contains all stellar material including stars during their lifetime as well as 
stellar remnants like neutron stars and black holes. Not included in $M_{\mathrm{Stars}}$ is the processed material which has been already returned to the ISM 
during the AGB and SN phase.

\subsection{Mass budget}

The TDG with filled IMFs consists of $8.35 \times 10^5\, \Msunrm$ of stars and $1.23 \times 10^8 \, \Msunrm$ of bound gas after 500 Myr. Therefore 63\% of the initial gas mass is still gravitationally bound to the TDG and acts as gas reservoir for further SF. The final stellar mass is 0.67\% of the total baryonic mass of $1.238 \times 10^8\,\Msunrm$.

The TDG with the truncated IMF contains $5.73 \times 10^6\, \Msunrm$ of stellar mass and $1.31 \times 10^8 \, \Msunrm$ of the gas mass (67 \% of the initial gas mass) is still gravitationally bound. Therefore the stellar mass that is present after 500 Myr is 4.2\% of the total baryonic mass of $1.37 \times 10^8\,\Msunrm$ at t = 500 Myr. 

Accounting also for the stars that have already died during the simulation, the total mass that was converted from gas into stars during the simulation time is $7.22\times10^6\,\Msunrm$ (3.7\% of the initial gas mass) for the truncated IMF and $1.17\times10^6\,\Msunrm$ (0.6\% of the initial gas mass) for the filled IMF run.

The time evolution of the bound gas mass and the SFR for both IMF cases are shown in Fig.~\ref{fig:MLsfh} (fourth panel and top panel, respectively). The most significant differences are found in the SFH, the total SF efficiency (0.6\% filled IMF, 3.7\% truncated IMF) and subsequently the stellar mass present at t = 500 Myr.  In spite of the large differences in the SFHs, the two considered models attain similar gas masses at the end of the simulation ($1.2 \times 10^8\,\Msunrm, 1.3 \times 10^8\,\Msunrm$). This implies that only between 33 and 37\% of the initial gas mass has been lost or converted into stars. Plumes of gas can be seen leaving the main body of the TDG in Fig.~\ref{fig:dens0500}, upper panels, but in both cases the inner ~2 kpc of the galaxy are still filled with gas and maintain roughly a spherical symmetry after 500 Myr of evolution. We can thus conclude that the feedback from SNe and stellar winds does not disrupt TDGs in spite of the lack of a dark matter halo. This confirms the conclusions reached in \citet{2007AA...470L...5R}.

\begin{figure}
\begin{center}
	\includegraphics[width=\linewidth]{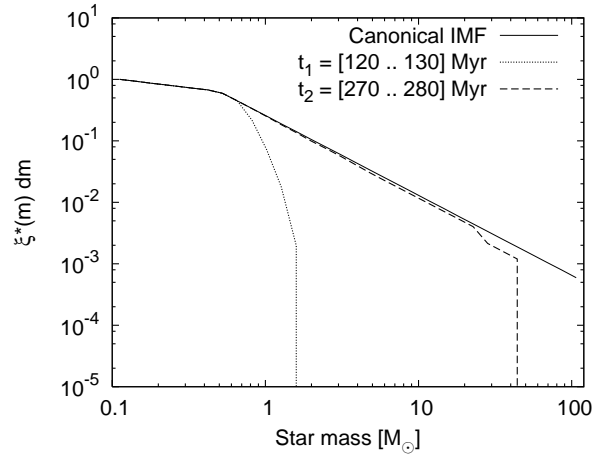}
\caption{The integrated galactic IMF for $t_1 = [120 \dots 130]\,\mathrm{Myr}$ (dotted line), $t_2 = [270 \dots 280]\,\mathrm{Myr}$ (dashed line) and the filled canonical IMF (solid line). }
\label{fig:igimf130}
\end{center}
\end{figure}

\subsection{Metallicity}

\citet{2009AA...499..711R} investigated the chemical evolution
within the IGIMF theory in terms of semi-analytical models.  A key
result of that investigation was that the global SFR strongly affects
the final [$\alpha$/Fe] ratios of galaxies.  The IGIMF theory predicts
in fact that galaxies with low SFRs have steeper IMF slopes at high
masses and lower upper-mass limits than galaxies with
higher SFRs.  This naturally leads to a reduced production of
$\alpha$-elements by massive stars and, consequently, to lower
[$\alpha$/Fe] ratios in dwarf galaxies, which are usually
characterized by low SFRs.  The resulting correlation between
[$\alpha$/Fe] ratio and galaxy masses was found to be in good
agreement with the available observations.  It is important to remark
that these results were obtained adopting spatially constant,
homogeneous SFRs in galaxies.  It was assumed that the global,
galactic-scale SFR solely determines the IGIMF.  However, the SF in a
galaxy is usually very inhomogeneously distributed. It is thus reasonable to 
expect that the IMF varies not only with time, but also with location within a galaxy.  
This approach was
used for instance by \citet{2008Natur.455..641P} to explain the
cut-off in H$\alpha$ radiation in the external regions of spiral
galaxies (where the SFRs are milder).  Observational evidence of the
variation of the IMF within galaxies is given by
\citet{2013MNRAS.428.3183D}.

\begin{figure}
\begin{center}
	\includegraphics[width=\linewidth]{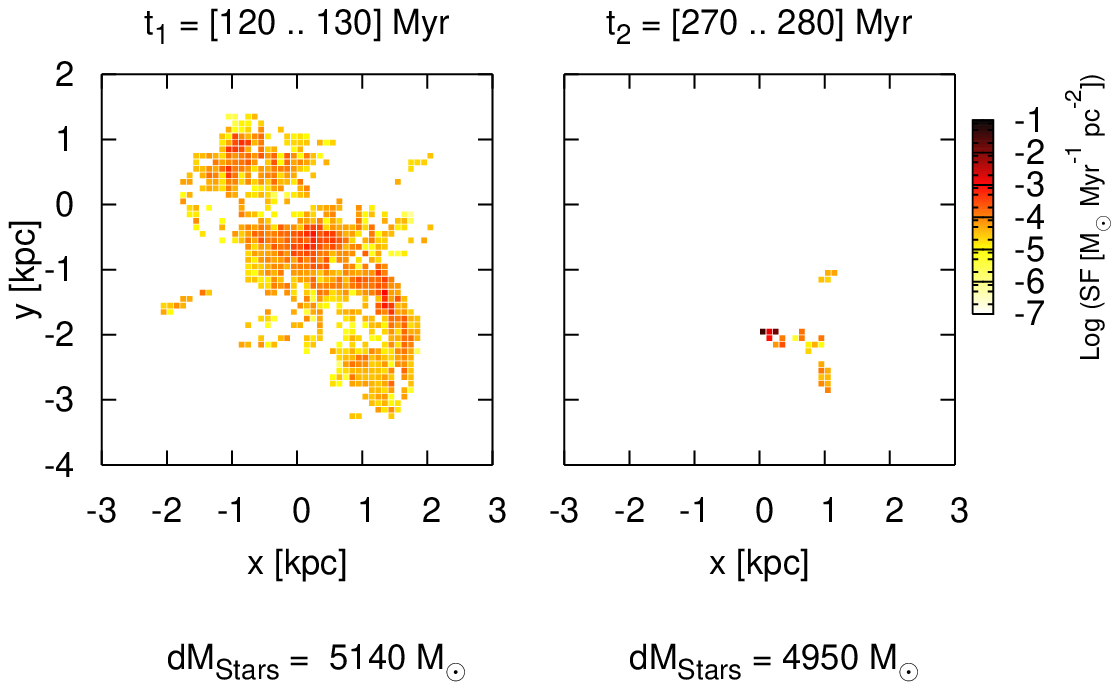}
	\includegraphics[width=\linewidth]{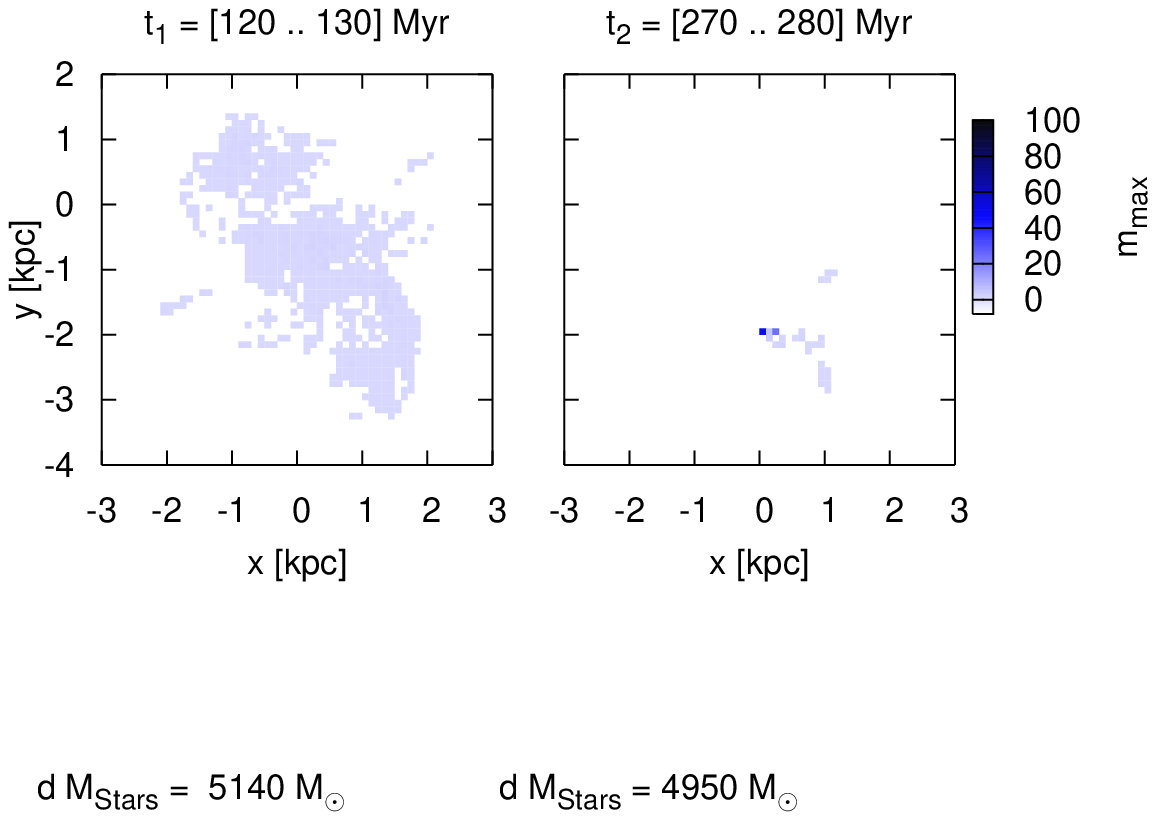}

\caption{2D star formation rate (upper panels) and maximal star masses (lower panels) for $t_1 = [120 \dots 130]\,\mathrm{Myr}$ and $t_2 = [270 \dots 280]\,\mathrm{Myr}$ for columns of $100\, \mathrm {pc} \times 100\, \mathrm {pc}$ in z-direction for the IGIMF model. The color coding in the upper figures represents the average SFR in 10 Myr within each column. The colors in the figures at the bottom show the maximal star mass (in $\Msunrm$) for each column. The IGIMF for $t_1$ and $t_2$ are plotted in Fig.~\ref{fig:igimf130}. Note that the TDG forms a star-burst cluster at (x,y) = (0,-2) kpc during $t_2$.}
\label{fig:sfnimf}
\end{center}
\end{figure}

\begin{figure}
	\includegraphics[width=\linewidth]{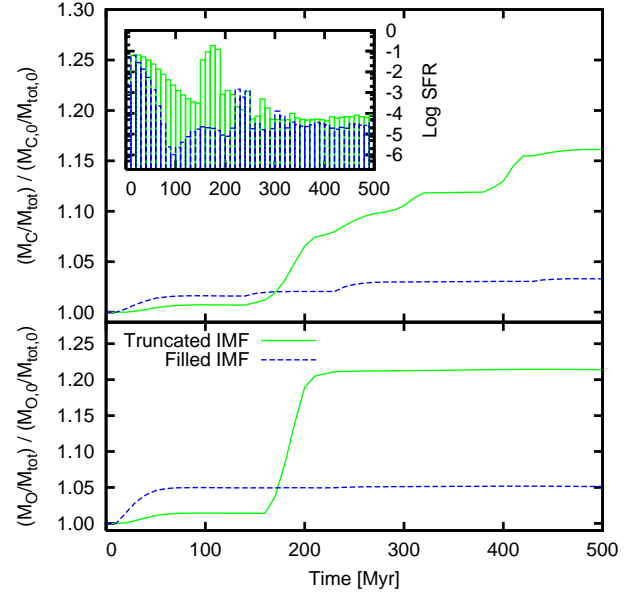}
	\caption{The evolution of the mass fractions of Carbon (upper panel) and Oxygen (lower panel) relative to the initial mass fraction for the simulated TDG with truncated (green solid line) and filled IMF (blue dashed line). The insert shows the SFR [$\Msunrm \, {\mathrm{yr}}^{-1}$] for both simulations.}
\label{fig:elmasses}
\end{figure}

The consequences for the chemical evolution of galaxies are obvious,
given the much larger number of SNeII for the truncated IMF model in
the time span $t_2$ as compared to those exploded in the time span
$t_1$.  These consequences can be quantified from 
Fig.~\ref{fig:elmasses}, which shows the evolution of the gas-phase
abundances of C and O for the truncated and filled IMF models.
For both models, there is an early phase of chemical enrichment (until
$t \approx$ 40 Myr), mainly due to the first SNeII.  Until $t \approx$ 150
Myr then, there is no further pollution of the ISM.  This is simply
due to the fact that the SFR is very low after $\approx$ 60 Myr for the
filled IMF model.  The SFR is instead non-negligible in the truncated
IMF model but, as we have seen, these localized, weak episodes of SF
do not lead to the formation of massive stars.  The production of
heavy elements (in particular O) by the massive stars formed in this
time interval is negligible.  After $t \approx$ 160 Myr a strong SF burst
occurs in the galaxy center and, this time, there are a large number of
SNeII polluting the ISM.  There is, consequently, a
sharp rise of O, mainly produced by SNeII on short time-scales.  A
significant fraction of C is produced by longer-living intermediate-mass stars. 
For this reason, the increase of the abundance of C after $\approx$ 160
Myr is less sharp, but still very evident.

It is important to point out that, at this stage, the mass of gas
restored to the ISM by dying stars and stellar winds is still a small
fraction (of the order of a few per cent) of the total galactic gas
content.  For this reason, the metal budget is still dominated by the
initial metallicity (0.3 Z$_\odot$) of the ISM and the increase in
abundance due to the chemical feedback is quite limited.

The difference in the gas metallicity between the simulations with truncated and filled IMFs after 500 Myr is displayed in the bottom 
panels of Fig.~\ref{fig:dens0500}. 

\subsection{Luminosity}

For every star cluster, the total stellar population mass $M_{\mathrm{spop}}$, age and $n_{\mathrm{imf,max}}$, the maximum IMF mass bin that is still populated by at least one star are stored. According to the number of stars in each (populated) mass bin, the total luminosity of this cluster can be calculated, assuming a mass-luminosity relation for individual main sequence stars of

\begin{eqnarray}
	\frac{L}{\Lsunrm} &=& 0.23 \left ( \frac{M}{\Msunrm} \right )^{2.3}	\quad 	(M \le 0.43\,\Msunrm) \, ,\\	 
	\frac{L}{\Lsunrm} &=& \left ( \frac{M}{\Msunrm} \right )^4	 	\quad 	(M>0.43\,\Msunrm) \, ,
\end{eqnarray}

\noindent
\citep[e.g.][]{2004adas.book.....D}. Integrating over all star particles in the galaxy, the total luminosity is derived given the underlying IMF. Fig.~\ref{fig:MLsfh} shows a medley of SFR, luminosity, stellar mass, and bound gas mass for TDGs with filled and truncated IMFs, respectively. The effect of the different IMF types is best seen in the beginning of the simulation where the SFRs are still comparable. Since there are massive stars in all stellar clusters for the filled IMF case, the total luminosity after the first SF events is more than an order of magnitude higher than for the case of the truncated IMF. At later stages, the difference in the total luminosity is dominated by very different dynamical evolutions and SFHs of the two simulations. For the truncated IMF the dependence on the spatial distribution of SF is highlighted again for comparing the SFRs and luminosities in the first 30 Myr (SFR $\approx 0.05\,\Msunrm\,\mathrm{yr}^{-1}$, L $\approx \, 7 \times 10^7 \,\Lsunrm$) and between 150 and 200 Myr (SFR $\approx 0.05 \dots 0.2\,\Msunrm\,\mathrm{yr}^{-1}$, L $\approx \, 2 \times 10^9 \,\Lsunrm$). As shown for $t_1$ and $t_2$ in Fig.~\ref{fig:sfnimf} in addition to the IGIMF also the total luminosity is dependent on whether there is low SF spread over a large volume or higher but more concentrated SF.

The total stellar masses deviate by a factor of 3 between the first 30 Myr and the starburst between 150 and 200 Myr, while the difference in the luminosity reaches 2 orders of magnitude (see second and middle panel of Fig.~\ref{fig:MLsfh}). Therefore the differences in the luminosity cannot be explained by the differences in the total stellar mass alone. For a more quantitative analysis the ratio of the total stellar masses have to be taken into account but here we only highlight the qualitative, significant difference. For both IMF simulations total mass to light ratios up to 10 can be reached at different stages throughout the simulation. As TDGs are not supposed to contain dark matter, the mass of the galaxy consists of stellar and gas mass only. 

\section{Discussion and Conclusion} \label{sec:discussion}

In order to explore the survival and evolution of TDGs we have performed hydrodynamical simulations of TDGs with different underlying IMFs and found drastic differences in the gas dynamics, star formation histories and chemical compositions. We studied the case where even for star clusters with masses as low as a few solar masses, the IMF is filled and therefore always a fraction of massive stars are produced. 

In a comparison run we account for the observed relation between SFR, enclosed cluster mass $M_{{\mathrm {ecl}}}$ and the maximal star mass 
$m_{{\mathrm {max}}}$ inside the star cluster and truncate the IMF at the high mass end where the IMF mass bin contains fewer than one star. 
Especially for dwarf galaxies with long episodes of low star formation rates the simulation resembles the observed IGIMF with a 
steeper slope than the IMF of individual clusters and a lack of very massive stars. We found that the IGIMF follows the descriptions from theoretical and semi-
analytical considerations very well during time intervals where the total SFR has a low variance within the galaxy. At times where the total SFR is dominated by 
a small volume with a very high SFR the IGIMF is dominated by this region as well.

For low or moderate SFRs the choice of how to treat the IMF, truncated or filled, determines the chemical and dynamical evolution in the galaxy. Massive stars have shorter lifetimes and therefore a star cluster where massive stars are present affects the gas phase faster. The additional energy input by stellar winds and radiation as well as the fractions of SNII  explosions provide enough feedback to prevent further SF in low SFR surroundings. The same region can continue SF for longer in the case of a truncated IMF, as the first SNeII happen later and also the wind and radiation feedback is dependent on the star mass, with fewer high mass OB stars injecting less stellar winds.

An IMF cutoff at high masses is motivated both by the observed $M_{\mathrm{ecl}}-m_{\mathrm max}$ relation (see Sec.~\ref{sec:imf} for details) as well as by the observed discrepancy between the SFRs derived from H$\alpha$ fluxes and those derived from FUV fluxes, which can be explained with truncated IMFs \citep{2009ApJ...706..599L}. In spite of their very different SFHs, TDGs with both truncated and filled IMFs self-regulate the SF already after about 300 Myr without getting disrupted. We conclude that DM-free objects with baryonic masses of dwarf galaxies in an external tidal field can survive an early star burst event independent of their IMF distribution. TDGs are even more resistant against stellar feedback processes when the IMFs of individual star clusters are truncated. This is especially interesting as in the TDG with truncated IMFs six times more mass is converted from gas into stars, compared to the TDG with filled IMFs. 

We show in Appendix~\ref{Sec:app} that a star cluster with $M_{\mathrm{tot}}= 500\,M_{\odot}$ releases almost twice the amount of energy when a fully-populated IMF is assumed than in the case of a truncated IMF. Therefore, truncated IMFs, in line with the IGIMF theory, increase the survivability of DM-free, low-mass objects, such as TDGs.

In an upcoming study (Ploeckinger et al., in prep.) we follow the evolution of a set of TDG with different initial metallicities and further investigate the conditions under which TDGs can survive their early evolution and turn into long-lived objects.

\section{Acknowledgements}
The authors are grateful for constructive suggestions by an anonymous referee that helped to improve the clarity of the publication. We want to thank P.~A. Duc for highly appreciated discussions and O.~Gnedin for helpful comments on the paper. The software used in this work was in part developed by the DOE NNSA-ASC OASCR Flash Center at the University of Chicago. S.P. was partly funded by within the DFG Priority Program ``Witnesses of Cosmic History: Formation and evolution of galaxies, their central black holes, and their environment" under the project no. HE1487/36  and the Austrian Science Fund FWF under project number P21097-N16. The numerical simulations are performed at the HPC astro-cluster of the Institute of Astronomy and at the Vienna Scientific Cluster (VSC-1, project- ID: 70128).

\bibliographystyle{mn2e}

\providecommand{\noopsort}[1]{}\providecommand{\singleletter}[1]{#1}

\appendix
\section{Truncated vs. filled IMF} \label {Sec:app}

The differences between truncated and filled IMFs are quantified in detail in this section.
We assume a star cluster where stars with a total stellar mass of $M_{\mathrm {tot}}$ are formed within the cluster formation time $\tau_{\mathrm{cl}}$. The range in star masses from $m_{\mathrm {imf, min}} = 0.1\,\Msunrm$ to $m_{\mathrm {imf, max}} = 120\,\Msunrm$ is logarithmically divided into 32 bins.

In general, the  total number of stars between $m_{\mathrm {imf, min}}$ and $m_{\mathrm {imf, max}}$ is given by:

\begin{eqnarray*}
	N_{\mathrm{tot}} &=& \int_{0.1 \Msunrm}^{120 \Msunrm} \xi(m) \mathrm{d}m 
\end{eqnarray*}

\noindent
For the function $\xi(m)$ by \citet{2001MNRAS.322..231K}, the total number of stars $N_{\mathrm{tot}}$ and the total mass of the star cluster $M_{\mathrm{tot}}$ are:

\begin{eqnarray*}
	N_{\mathrm{tot}} &=&c_1 \int_{m_{\mathrm {imf, min}} }^{0.5 \Msunrm} m^{-1.3} \mathrm{d}m + c_2 \int_{0.5 \Msunrm}^{m_{\mathrm {imf, max}} } m^{-2.3} \mathrm{d}m  \\
	M_{\mathrm{tot}} &=&c_1 \int_{m_{\mathrm {imf, min}} }^{0.5 \Msunrm} m^{-0.3} \mathrm{d}m + c_2 \int_{0.5 \Msunrm}^{m_{\mathrm {imf, max}} } m^{-1.3} \mathrm{d}m 
\end{eqnarray*}

\noindent
For the condition that $\xi(m)$ is smooth at $m=0.5\,\Msunrm$, the constant $c_1 = 2 \cdot c_2$ and for a given total cluster mass $M_{\mathrm{tot}}$, both constants are determined.

\noindent
In order to highlight the differences in the stellar feedback between a star cluster with truncated and filled IMF, we calculate the energy and metal feedback by massive stars from a star clusters with 500 $\Msunrm$. We use for the discretization the same 32 logarithmical mass bins as in the simulation.

\subsection{Truncated IMF ($M_{\mathrm {tot}} = 500 \, \Msunrm$):}
As illustrated in Fig.~\ref{fig:imf}, for the chosen mass binning the last mass bin, that is populated with at least one star is the 22nd mass bin ranging up to a maximal star mass of $m_{\mathrm{t, 500 }\Msunrm \mathrm{, max}} = 13.1\, \Msunrm$. For the truncated IMF the total mass is not distributed over the whole mass range from $m_{\mathrm{min}}$ to $m_{\mathrm{max}}$ but only up to the upper end of the last mass bin $m_{\mathrm{max, 500}\,\Msunrm}$. The constant $c_{t,500\,\Msunrm}$ for the truncated IMF of a 500 $\Msunrm$ star cluster is calculated by

{\scriptsize
\begin{eqnarray*}
	M_{\mathrm{tot}} =  c_{t,500\,\Msunrm}  \left ( 2  \int_{m_{\mathrm{imf, min}}}^{0.5 \Msunrm} m^{-0.3} \mathrm{d}m  +  \int_{0.5 \Msunrm}^{m_{\mathrm{t, 500 }\Msunrm \mathrm{, max}} } m^{-1.3} \mathrm{d}m \right ) \, .
\end{eqnarray*}  }

\noindent
The total number of stars in the mass bins with average masses $\bar{m}(i)= M(i) / N(i)$ of more than $8\, \Msunrm$ and therefore the number of SNII explosions in this star cluster is calculated with

\begin{eqnarray*}
	N_{\mathrm{SNII},t,500\,\Msunrm} &=& c_{t,500\,\Msunrm} \int_{m_{\mathrm{SNII, min, bin}}}^{m_{\mathrm{t, 500 }\Msunrm \mathrm{, max}} } m^{-2.3} \mathrm{d}m \, ,
\end{eqnarray*}

\noindent
where $m_{\mathrm{SNII, min, bin}} = 8.4\,\Msunrm$ for the chosen mass bin discretization. This results in a total number of $N_{\mathrm{SNII},t,500\,\Msunrm} = 2.8$ massive stars with a total mass of $M_{\mathrm{SNII},t,500\,\Msunrm} = 29.2 \,\Msunrm$ in the mass range between $m_{\mathrm{SNII, min, bin}} = 8.4\,\Msunrm$ and $m_{\mathrm{t, 500 }\Msunrm \mathrm{, max}}  = 13.1\,\Msunrm$.

\subsection{Filled IMF ($M_{\mathrm {tot}} = 500 \, \Msunrm$):}

For the fully populated IMF, where $m_{\mathrm {imf, max}} = 120 \, \Msunrm$ in every case, the number of massive stars ($m>8\, \Msunrm$) is given by

{\scriptsize
\begin{eqnarray*}
	 c_{f,500\,\Msunrm} = M_{\mathrm{tot}} \left ( 2 \int_{m_{\mathrm{imf, min}}}^{0.5 \Msunrm} m^{-0.3} \mathrm{d}m + \int_{0.5 \Msunrm}^{m_{\mathrm{imf, max}}} m^{-1.3} \mathrm{d}m \right )^{-1}  \\
\end{eqnarray*} }

\noindent and 

\begin{eqnarray*}
	N_{\mathrm{SNII},f,500\,\Msunrm} = c_{f,500\,\Msunrm} \int_{m_{\mathrm{SNII, imf, bin}}}^{m_{\mathrm{imf, max}}} m^{-2.3} \mathrm{d}m \, .
\end{eqnarray*}

\noindent
This results in $N_{\mathrm{SNII},f,500\,\Msunrm} =  5.2$ massive stars with a total mass of $M_{\mathrm{SNII},f,500\,\Msunrm} = 107.48 \,\Msunrm$ in the mass range from $m_{\mathrm{SNII, imf, bin}}= 8.4\,\Msunrm$ to $m_{\mathrm{imf, max}}=120\,\Msunrm$. In this case, the whole IMF is populated, but the number of stars in high-mass bins is less than 1. For example, the star cluster consists of 0.057 stars with an average mass of 107 $\Msunrm$, 0.075 stars with 85.79 $\Msunrm$ and 0.1 stars of 68.74 $\Msunrm$.

\subsection{Stochastic IMF} \label{app:stochastic}

 We stochastically filled a set of $10^4$ \citet{2001MNRAS.322..231K} IMFs for a $500\,\Msunrm$ test cluster. For each of the IMFs, the number and total mass of massive stars ($M \ge 8\,\Msunrm$) as well as the total yields are calculated. For a better comparison, the same mass binning (32 logarithmic bins) as for the truncated and fully-populated IMFs is used. The $500 \,\Msunrm$ cluster contains in this case on average $N_{\mathrm{SNII},s,500\,\Msunrm} =  5.1 \pm 1.9$ SNII progenitor stars with a total mass of $M_{\mathrm{SNII},s,500\,\Msunrm} = (102 \pm 48) \, \Msunrm$. 

The average number of SNII events as well as the average total mass of massive stars in the test cluster are close to the values for the fully-populated IMF, although the deviation of individual star clusters can be large. An overview of all values for the three different IMF treatment can be found in Tab.~\ref{tab:enercomp} with the total yields listed in Tab.~\ref{tab:yieldcomp}.

\subsection{Energy feedback:}
Compared to the truncated IMF, the fully populated IMF has not only almost twice as many SNeII explosions, but also the delay between
the cluster formation and the SNeII events are very different. The lifetime of a 120 $\Msunrm$ star is, depending on its metallicity, in the range of [$3.11 \dots 3.32$] Myr, while for an 8 $\Msunrm$ star the lifetime is around 40 Myr \citep{1998AA...334..505P}.
For a star cluster of 500 $\Msunrm$ and a truncated IMF, the energy feedback by SNeII has a delay of 40 Myr, where the energy of at least one SNII is injected into the ISM. The same star cluster with a fully-populated IMF injects the energy from fractions of SNeII already 
from about 3 Myr onwards.

Although the SF is regulated by stellar wind feedback in both cases, the additional fractions of SNeII in the filled IMF case, very early after the formation of the cluster, over-regulates further SF compared to clusters with a truncated IMF. 

\begin{table} 
\caption{Overview of the number $N_{\mathrm{SNII}}$ of SNeII events and the total mass $M_{\mathrm{SNII}}$ of SNII progenitor stars (stars in the mass bins with an average mass of $\ge\,8\,\Msunrm$) in the $500\,\Msunrm$ test cluster for different IMF realizations.}
\begin{center}
\begin{tabular}{l|c|c|c}
\hline
	& Truncated IMF& Stochastic IMF & Full IMF \\
\hline
$N_{\mathrm{SNII}}$ 			& 2.8	& 5.1 $\pm$ 1.9 	&  5.2 \\
$M_{\mathrm{SNII}}\, \Msunrm$ 	& 29.2 	& 102 $\pm$ 48	&  107.5 
\end{tabular}
\end{center}
\label{tab:enercomp}
\end{table}

\subsection{Metal enrichment of the ISM:}

The stellar nucleosynthesis and, as a result, the stellar yields are dependent on the initial mass and metallicity of the progenitor star. Therefore, the composition and the total masses of the released elements will be different whether one 100 $\Msunrm$ star or ten 10 $\Msunrm$ stars inject their metals to the ISM. For the example star cluster with $500 \, \Msunrm$, not only the total number of massive stars differs between the truncated IMF and the filled IMF case, but also the metal enrichment of the ISM. The differences are quantified in Tab.~\ref{tab:yieldcomp} for an initial metallicity of Z = 0.004. The total stellar yields are calculated for the average mass of every mass bin with the yields from \citet{1998AA...334..505P}.

For the example cluster of $500 \, \Msunrm$, the total mass of massive stars is $M_{\mathrm{SNII},t,500\,\Msunrm} = 29.2 \,\Msunrm$ for the truncated IMF and $M_{\mathrm{SNII},f,500\,\Msunrm} = 107.48 \,\Msunrm$ for the filled IMF. If the stellar yields for massive stars only scale with the total mass, $Y_{\mathrm{full}}/Y_{\mathrm{trunc}}$ would be around 3.7 for every element. The last column of Tab.~\ref{tab:yieldcomp} shows the overabundance of all elements, especially O, Ne and Mg in the star cluster with the filled IMF. Only Fe is returned to the ISM in comparable masses for both test star clusters leading to a relative under-abundance of Fe in contrast to the other elements. The same star cluster with a truncated IMF contributes significantly less to the metal enrichment of the ISM, especially in O, Ne, and Mg.

\begin{table}
\caption{Total stellar yields \citep{1998AA...334..505P} of stars in mass bins with $\bar{m} > 8\,\Msunrm$ for a star cluster with $M_{\mathrm{tot}}=500\,\Msunrm$ and an initial metallicity of Z=0.004. Yields are listed for a truncated, a fully-populated and a stochastic IMF. In addition the ratios between the full and the truncated IMFs for this test cluster are listed in the column $Y_{\mathrm{full}}/Y_{\mathrm{trunc}}$.}
\begin{center}
\begin{tabular}{l|c|c|c|c}
\hline
	& $Y_{\mathrm{trunc}}$ [$\Msunrm$] & $Y_{\mathrm{full}}$ [$\Msunrm$]	&	$Y_{\mathrm{full}}/Y_{\mathrm{trunc}}$ & $Y_{\mathrm{stoch}}$ [$\Msunrm$]\\
\hline
H 	&13.99	&42.31	&3.02	&	$40	\pm 		16$\\
He	&8.700	&29.37	&3.38	&	$28	\pm 		13$\\
C 	&0.358	&1.117	&3.12	&	$1.07	\pm 		0.42$\\
N 	&0.028	&0.114	&4.03	&	$0.107	\pm 		0.06$\\
O 	&1.151	&8.082	&7.02	&	$7.7		\pm 		3.7$\\
Ne 	&0.179	&1.045	&5.85	&	$0.99	\pm 		0.55$\\
Mg 	&0.060	&0.378	&6.29	&	$0.36	\pm 		0.20$\\
Si 	&0.175	&0.607	&3.47	&	$0.59	\pm 		0.32$\\
S 	&0.081	&0.303	&3.72	&	$0.29	\pm 		0.17$\\
Ca 	&0.015	&0.040	&2.75	&	$0.039	\pm 		0.02$\\
Fe 	&0.477	&0.669	&1.40	&	$0.66	\pm 		0.29$\\
X 	&0.013	&0.048	&3.82	&	$0.045	\pm 		0.03$\\
\hline
\end{tabular}
\end{center}
\label{tab:yieldcomp}
\end{table}%

\end{document}